\documentclass[journal]{IEEEtran}
\IEEEoverridecommandlockouts

\usepackage{graphicx}
\usepackage{subfigure}
\usepackage{amsmath,amsthm,amsfonts,amssymb}
\usepackage{cite}
\usepackage{bm}
\usepackage{bbm}
\usepackage{url}
\usepackage{array}
\usepackage{color}
\usepackage{multirow}
\usepackage{booktabs}
\usepackage[table,xcdraw]{xcolor}
\usepackage{enumerate}
\usepackage{epstopdf}
\usepackage{tabularx}
\usepackage{algorithm}
\usepackage[english]{babel}
\usepackage{algpseudocode}

\newtheorem{thm}{\textbf{Theorem}}
\newtheorem{cor}{Corollary}

\newtheorem{lem}{\textbf{Lemma}}
\newtheorem{rem}{Remark}

\newtheorem{defi}{\textbf{Definition}}

\newenvironment{NewProof}{{\noindent\it Proof.}}{\hfill $\blacksquare$\par}

\algrenewcommand{\algorithmiccomment}[1]{\hfill// #1}

\allowdisplaybreaks[4]

\begin{document}
\title{Efficient FFT Computation in IFDMA Transceivers}

\author{
Yuyang~Du,~\IEEEmembership{Student Member,~IEEE},
Soung~Chang~Liew,~\IEEEmembership{Fellow,~IEEE},
Yulin~Shao,~\IEEEmembership{Member,~IEEE}
\thanks{Y. Du and S. C. Liew are with the Department of Information Engineering, The Chinese University of Hong Kong, Shatin, New Territories, Hong Kong SAR (e-mail: \{dy020, soung\}@ie.cuhk.edu.hk).}
\thanks{Y. Shao is with the Department of Electrical and Electronic Engineering, Imperial College London, London SW7 2AZ, U.K. (e-mail: y.shao@imperial.ac.uk).}
}

\maketitle

\begin{abstract}
Interleaved Frequency Division Multiple Access (IFDMA) has the salient advantage of lower Peak-to-Average Power Ratio (PAPR) than its competitors like Orthogonal FDMA (OFDMA). A recent research effort put forth a new IFDMA transceiver design significantly less complex than conventional IFDMA transceivers. The new IFDMA transceiver design reduces the complexity by exploiting a certain correspondence between the IFDMA signal processing and the Cooley-Tukey IFFT/FFT algorithmic structure so that IFDMA streams can be inserted/extracted at different stages of an IFFT/FFT module according to the sizes of the streams. Although the prior work has laid down the theoretical foundation for the new IFDMA transceiver's structure, the practical realization of the transceiver on specific hardware with resource constraints has not been carefully investigated. This paper is an attempt to fill the gap. Specifically, this paper puts forth a heuristic algorithm called multi-priority scheduling (MPS) to schedule the execution of the butterfly computations in the IFDMA transceiver with the constraint of a limited number of hardware processors. The resulting FFT computation, referred to as MPS-FFT, has a much lower computation time than conventional FFT computation when applied to the IFDMA signal processing. Importantly, we derive a lower bound for the optimal IFDMA FFT computation time to benchmark MPS-FFT. Our experimental results indicate that when the number of hardware processors is a power of two: 1) MPS-FFT has near-optimal computation time; 2) MPS-FFT incurs less than 44.13\% of the computation time of the conventional pipelined FFT.
\end{abstract}


\section{Introduction}\label{sec-I}
Interleaved Frequency Division Multiple Access (IFDMA) is a broadband signal modulation and multiple-access technology for advanced wireless communications systems. Compared with other similar technologies \cite{Ref_B1,Ref_B2}, such as Orthogonal FDMA (OFDMA), a salient feature of IFDMA is that its signal has a low Peak-to-Average Power Ratio (PAPR) \cite{Ref_B3}. The low PAPR of IFDMA brings about low signal distortion and high power efficiency when amplified by a power amplifier \cite{ref_D5}, making it a promising ``green technology" for future wireless communications systems in which performance and energy efficiency are a concern \cite{Ref_B4}. For the widespread deployment of IFDMA, however, the complexity of IFDMA transceivers needs to be minimized.

Recent works in \cite{ref_1,ref_2} put forth a new class of IFDMA transceivers that is significantly less complex than conventional IFDMA transceivers. In essence, the design in \cite{ref_1,ref_2} allows a single IFFT/FFT module to perform the multiplexing/demultiplexing of multiple IFDMA data streams of different sizes, obviating the need to have IFFT/FFT modules of different sizes to cater to different IFDMA data streams. The key is to insert (for multiplexing at the transmitter) and extract (for demultiplexing at the receiver) the IFDMA data streams at different stages of the IFFT/FFT module according to their sizes.

In this paper, we refer to the IFDMA transceiver design in \cite{ref_1,ref_2} as ``compact IFDMA" to capture that it performs IFDMA functions in a compact and efficient manner. We refer to the specially designed IFFT/FFT in compact IFDMA as ``compact IFDMA IFFT/FFT".

Although \cite{ref_1,ref_2} have laid down the theoretical foundation for compact IFDMA, its implementation in specific hardware (e.g., ASIC and FPGA) under resource constraints presents several technical challenges yet to be addressed. Take FPGA for example. Each butterfly computation unit in the IFFT/FFT can be implemented as one processor unit in an FPGA. Due to the resource constraints, we can implement a limited number of such processor units in the FPGA. Consequently, the amount of parallelism that can be achieved is limited, and scheduling the butterfly computation processes in compact IFDMA IFFT/FFT to minimize the computation latency is an issue.

Although mature hardware implementations of conventional IFFT/FFT are available \cite{ref_B5,Ref_B6,Ref_B7}, they are not well matched to the compact IFDMA IFFT/FFT structure. As will be verified by the experimental results in Section \ref{sec-V}, conventional IFFT/FFT implementations incur a much longer computation latency than our specially designed compact IFDMA IFFT/FFT.

To achieve low latency, our IFDMA IFFT/FFT implementation schedules the butterfly computations within an IFFT/FFT network by taking into account 1) the precedence relationships among the butterfly computations; and 2) the fact that not all butterfly computations within a full IFFT/FFT network need to be executed in the multiplexing and demultiplexing of IFDMA data streams.

This paper focuses on the design of compact IFDMA FFT and skips the design of compact IFDMA IFFT. The Cooley-Tukey FFT and IFFT have similar decomposition structures, and the design of compact IFDMA IFFT is similar to that of IFDMA FFT. Interested readers are referred to Section IV and V of \cite{ref_1} for details on the slight differences between compact IFDMA FFT and IFFT.

\subsection{Quick Review and Motivation}
\begin{figure}[htbp]
  \centering
  \includegraphics[width=3.0in]{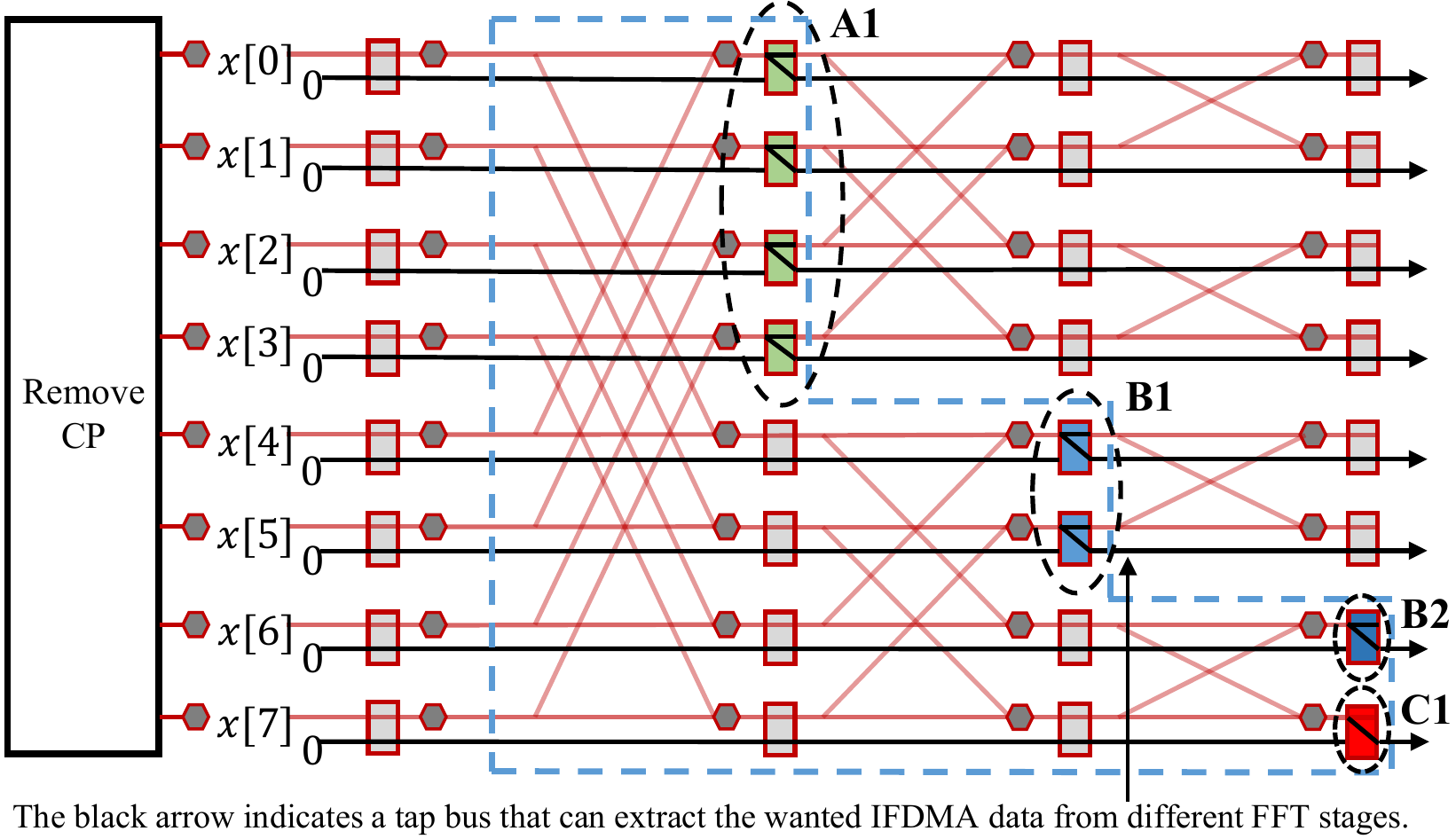}\\
  \caption{An example on how the receiver extracts/demultiplexes the IFDMA data streams embedded in IFDMA signals.}
\label{fig:1}
\end{figure}

We first briefly review how the compact IFDMA receiver works. Consider a scenario in which there are three users, A, B, and C, in an eight-subcarrier IFDMA system. Users A, B, and C require four subcarriers, three subcarriers, and one subcarrier, respectively. With the resource allocation scheme expounded in \cite{ref_2}, a four-subcarrier IFDMA stream A1 is assigned to user A; a two-subcarrier IFDMA stream B1 plus a single-subcarrier IFDMA stream B2 are assigned to user B; and a single-subcarrier IFDMA stream C1 is assigned to user C. Fig. \ref{fig:1} shows the receiver's FFT \cite{ref_1} responsible for extracting/demultiplexing the respective IFDMA streams from the IFDMA signals.

As illustrated in Fig. \ref{fig:1}, not all butterfly computations need to be executed. For example, data stream A1 can be extracted after the first-stage butterfly computations associated with it. The required data are extracted by four tap buses, bypassing the subsequent butterfly computations. Black arrows indicate the tap busses in Fig. \ref{fig:1} (details on the tap bus design can be found in Fig. 7 of \cite{ref_1}). Besides A1, Fig. \ref{fig:1} also shows the early extraction of B1. In general, only the IFDMA streams with one subcarrier need to go through the full FFT operations from the first stage to the last stage. All the butterfly computations outside the blue dashed line in Fig. \ref{fig:1} can be omitted.

\begin{figure}[htbp]
  \centering
  \includegraphics[width=3.0in]{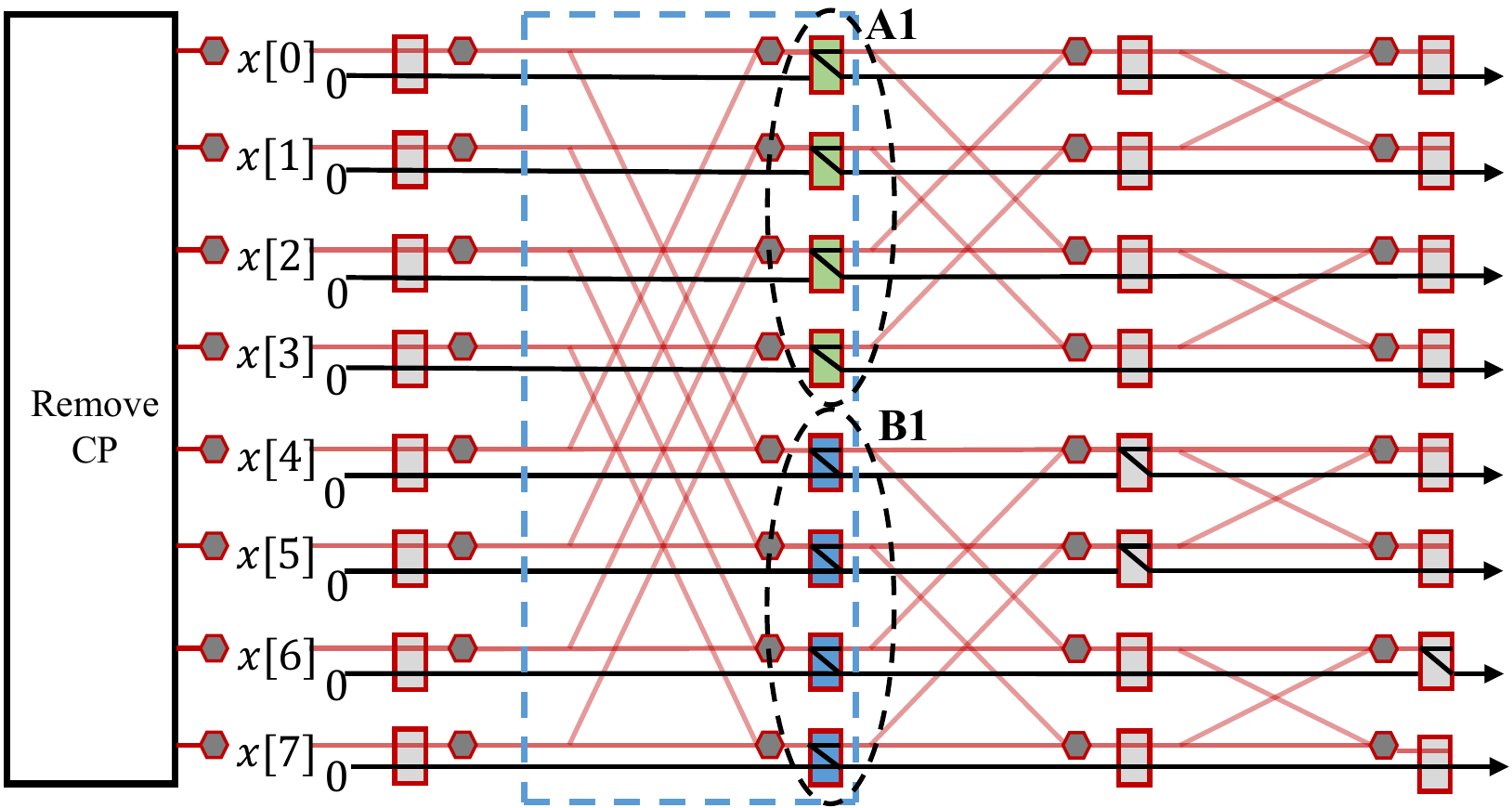}\\
  \caption{An example on a compact IFDMA FFT network with two users.}
\label{fig:2}
\end{figure}

In general, the butterfly computations that can be omitted depend on the users' subcarrier allocation. Fig. \ref{fig:2} shows another example in which two users, A and B, are each allocated four subcarriers. During the operation of an IFDMA system, users may come and go, and their subcarrier demands may vary from time to time. This calls for a quick scheduling algorithm to determine the order of the butterfly computations in accordance with the active IFDMA subcarrier allocation at any moment in time, with the target of minimizing the computation latency.

Conventional FFT implementations, however, are too rigid to serve this purpose. Three conventional FFT implementations are the serial FFT implementation \cite{ref_B5}, the pipelined FFT implementation \cite{Ref_B6}, and the parallel FFT implementation \cite{Ref_B7}. In these implementations, butterfly computations are executed in a fixed manner, i.e., from left to right, and from top to bottom. This rigidity is not a critical issue if the full FFT needs to be computed and the desired data can only be obtained after the final FFT stage. For compact IFDMA FFT, using the schedules of such conventional FFT implementations is not optimal. In Fig. \ref{fig:1}, for example, a conventional FFT implementation will execute the five unnecessary butterfly computations outside the dashed line.

This paper centers on a flexible computation scheduling algorithm tailored for IFDMA FFT. In general, task scheduling refers to the scheduling of a set of tasks with precedence relationships among them. A task cannot proceed before the inputs of the task are available, and the inputs of one task may be outputs of some preceding tasks. If we treat each butterfly computation as a task, it is then obvious that there are precedence relationships among the butterfly computations in the structure of the FFT. The precedence relationships can be captured by a directed acyclic graph (DAG) in which the vertexes are the tasks, and the directed edges indicate the precedence relationships between tasks. We refer to this special precedence graph as the FFT precedence graph.

In FPGA implementation, for example, we could design a customized butterfly computation processor and have a fixed number of such processors within the FPGA. The goal of the scheduling algorithm is to find an execution order for the tasks that can minimize the overall IFDMA FFT computation, subject to constraints of the precedence relationships and the fixed number of processors.

The general scheduling problem is NP-complete. However, we will provide evidence that our scheduling algorithm for IFDMA FFT is near-optimal. The algorithm 1) works well with arbitrary numbers of processors; 2) works well for random subcarrier allocations; and 3) has much lower computation time than conventional FFT implementations.

\subsection{Related Works}
Task scheduling subject to precedence constraints is tough in general. Karp first studied this problem in his seminal paper \cite{Ref_A1}, where the task scheduling problem is reduced to a 3-satisfiability (3SAT) problem and proved to be one of the first 21 NP-complete problems. Subsequently, \cite{Ref_A2} found that task scheduling to minimize the computation time subject to a fixed number of processors and a uniform task processing time is also NP-complete.

Our scheduling problem, unlike the general scheduling problem, is specific to the FFT precedence graph. As far as task scheduling subject to the FFT precedence graph is concerned, there have been two main research focuses.

The first line of research focused on memory access control. For example, \cite{Ref_A3} studied the scheduling of butterfly tasks to minimize the required data transfers between memories of different hierarchies within a device (e.g., between registers and cache). For compact IFDMA, every output of a butterfly task can potentially be the wanted data of an IFDMA stream, due to changing subcarrier allocations. For example, the two subcarrier allocations in Fig. \ref{fig:1} and Fig. \ref{fig:2} result in IFDMA streams being extracted at different places. For flexible IFDMA stream extraction, our implementation writes the outputs of all butterfly computations into a buffer immediately after they are generated. A subsequent signal processing module, such as a decoder after FFT, can collect the data stream and begin the next-step signal processing as soon as the data is available. In Fig. \ref{fig:1}, for example, a decoder can obtain the data stream A1 immediately after the first-stage butterfly tasks. In our FFT implementation, since all outputs of the butterfly computations are written into the memory, the optimization of memory access control is not our concern.

The second line of research focused on task scheduling within the conventional FFT implementations to speed up computation over specialized hardware. For example, \cite{Ref_A5} improved the butterfly task scheduling in a very-long-instruction-word (VLIW) digital signal processors (DSP) chip using a software pipelining technique called modulo scheduling. This scheduling algorithm exploits the instruction-level parallelism (ILP) feature in the VLIW DSP platform to schedule multiple loop iterations in an overlapping manner \cite{Ref_A10}. In \cite{Ref_A6,Ref_A7}, the authors put forth the celebrated Fastest Fourier Transform in the West (FFTW) C library for the FFT implementation on multiprocessor computers. Other works like UHFFT \cite{Ref_A8} and SPIRAL \cite{Ref_A9} also contributed to the FFT implementation on multiprocessor computers and DSP chips. What distinguishes our implementation from the above works is that our design omits the unnecessary butterfly computations in the IFDMA FFT. To our best knowledge, the task-scheduling problem in a partial FFT network has not been studied in the literature.

\subsection{Contributions and Findings}
We summarize the contributions of this paper in the following. We put forth an algorithm, referred to as multi-priority scheduling (MPS), to find a near-optimal schedule for compact IFDMA FFT computation in terms of computation latency. To establish the near optimality of MPS, we derive a lower bound for the optimal computation latency and show that MPS can find a schedule with computation latency approaching the lower bound.

We refer to the FFT computation scheduled by MPS as MPS-FFT. We conducted large-scale experiments to comprehensively study the performance of MPS-FFT with up to 1024 subcarriers. Two main experimental results are as follows:
\begin{enumerate}[1)]
\item When the number of processors is power-of-two, the computation time of MPS-FFT can reach the lower bound with a probability approaching one. Quantitatively, the probability of reaching the lower bound is larger than $\sqrt[{{\xi }_{0}}+1]{0.05}$, where ${{\xi }_{0}}$ is more than 2.5 million in our experiment.
\item When the number of processors is arbitrary, the MPS-FFT's computation time can reach the lower bound with a probability of at least 98.70\%. In the few cases where the MPS-FFT computation time fails to reach the lower bound, it is no more than 1.06 times the lower bound.
\end{enumerate}

Our experiments also demonstrate that MPS-FFT has a much smaller computation time than the conventional FFT schemes do. The speedup achieved by MPS-FFT can be attributed to two factors: (i) MPS-FFT can omit at least 11.21\% of the butterfly tasks thanks to the structure of the compact IFDMA (ii) MPS-FFT has a more than 98\% utilization rate of the butterfly processors when the number of processors is a power of two, almost twice higher than conventional pipelined FFT implementations with the same number of processors. Thanks to the above two benefits, the computation time of MPS-FFT with power-of-two processors is less than 44.13\% of that of the conventional pipelined FFT with the same number of processors.

\section{Problem Formulation and Definitions}\label{sec-II}
\subsection {Problem Formulation}
We consider an $N$-point FFT in a compact IFDMA receiver. In the FFT network, there are $n={{\log }_{2}}N$ FFT stages, with each stage having ${{2}^{n-1}}$ butterfly tasks.

For IFDMA subcarrier allocation, it is convenient to first map subcarriers to ``bins" through a bit-reversal mapping so that the subcarrier indexes, after bit reversal, become the bin indexes, and subcarrier allocation becomes bin allocation (see Section III of \cite{ref_2} on bit-reverse index mapping of subcarriers to bins). This mapping process allows an IFDMA stream to be allocated consecutive bins that correspond to regularly-interspersed subcarriers, as is needed for the IFDMA stream.

\begin{figure}[htbp]
  \centering
  \includegraphics[width=3.3in]{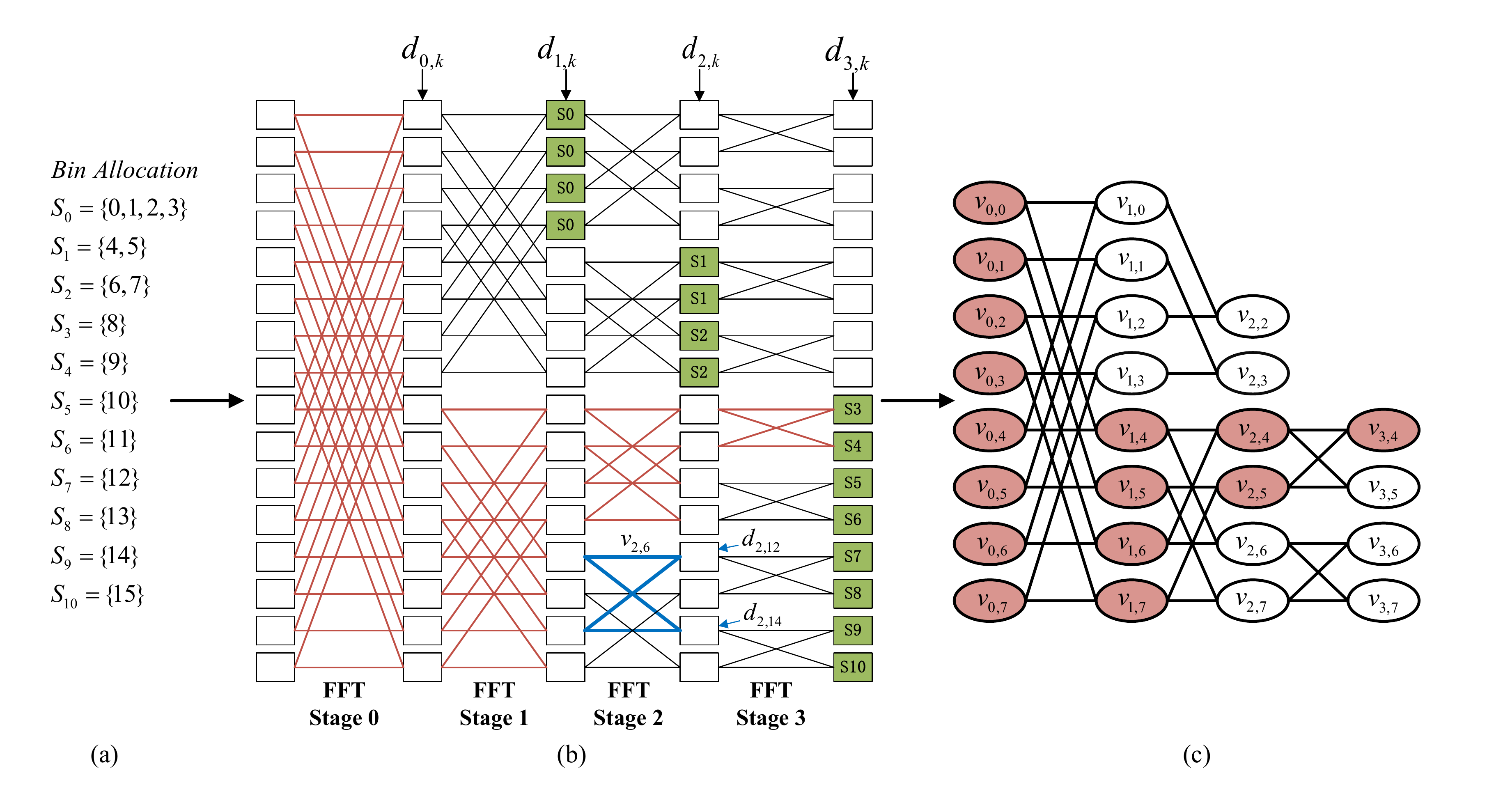}\\
  \caption{An example to illustrate how to obtain an FFT precedence graph given $\{{{S}_{0}},{{S}_{1}},...{{S}_{R-1}}\}$.}
\label{fig:3}
\end{figure}

A bin allocation can be expressed by an ordered list $\{{{S}_{0}},{{S}_{1}},...{{S}_{R-1}}\}$, where $R$ is the number of IFDMA streams, and ${{S}_{r}}\left( r=0,...,R-1 \right)$ is the set of bins allocated to IFDMA stream $r$. Fig. \ref{fig:3} gives an example of the construction of the FFT precedence graph given an IFDMA bin allocation. Fig. \ref{fig:3}(a) shows a specific bin allocation for a system with 16 subcarriers and 11 IFDMA streams. IFDMA stream 0 is allocated the bins ${{S}_{0}}=\{0,1,2,3\}$, IFDMA stream 1 is allocated the bins ${{S}_{1}}=\{4,5\}$, and so on. Fig. \ref{fig:3}(b) shows the corresponding IFDMA FFT network for the receiver, where the green boxes are where the data of the IFDMA streams can be extracted. Fig. \ref{fig:3}(c) shows the corresponding FFT precedence graph. A vertex in the precedence graph corresponds to a $2\times 2$ butterfly computation in the FFT network.

We next explain how the precedence graph in Fig. \ref{fig:3}(c) is obtained. Consider stage $i \text{ }(i=0,1,...,\log_2 N-1)$ of the FFT network. We denote task $j\text{ }(j=0,1,...,N/2-1)$ and FFT output $k\text{ }(k=0,1,...N-1)$ at stage $i$ by ${{v}_{i,j}}$ and ${{d}_{i,k}}$, respectively. With respect to Fig. \ref{fig:3}(b), the butterfly computation in blue corresponds to task 6 (enumerating from top to bottom) at stage 2 (enumerating from left to right). Thus, the task is written as ${{v}_{2,6}}$, and the outputs of this task are written as ${{d}_{2,12}}$ and ${{d}_{2,14}}$. The index $k$ in ${{d}_{i,k}}$ is labeled from top to bottom, with $k=0$ corresponding to a rectangular box at the top and $k=15$ corresponding to a rectangular box at the bottom.

Let $D$ denote the set of desired FFT outputs of all IFDMA streams, i.e., green boxes in Fig. \ref{fig:3}(b). Further, let $V$ denote the set of tasks that need to be executed to obtain these desired FFT outputs. The precedence graph in Fig. \ref{fig:3}(c) is obtained as follows.

In Fig. \ref{fig:3}(b), to obtain output ${{d}_{3,8}}$ required by IFDMA stream ${{S}_{3}}$, task ${{v}_{3,4}}$ needs to be completed. However, before task ${{v}_{3,4}}$ can be executed, tasks ${{v}_{2,4}}$ and ${{v}_{2,5}}$ need to be executed first, because their outputs ${{d}_{2,8}}$ and ${{d}_{2,9}}$ are inputs to ${{v}_{3,4}}$. Further, before ${{v}_{2,4}}$ and ${{v}_{2,5}}$ can be executed, some tasks in the prior stage need to be executed, and so on and so forth. In Fig. \ref{fig:3}(b), we mark all of the butterflies that need to be executed to obtain ${{d}_{3,8}}$ in pink. The vertexes in Fig. \ref{fig:3}(c) that correspond to these butterfly tasks are also pink. By tracing backward from the desired IFDMA stream outputs, we can obtain all the tasks that need to be executed and the precedence relationships between them to arrive at the overall precedence graph.

We note that every task except those at stage 0 requires the outputs of two specific tasks in the prior stage as its inputs, and the task cannot be executed unless these two preceding tasks are completed first. We define the concept of parent vertex and child vertex as follows:
\begin{defi}[Parent Vertex and Child Vertex]\label{defi:1}
If ${{v}_{i+1,{{j}_{0}}}}$ cannot be executed unless ${{v}_{i,{{j}_{1}}}}$ and ${{v}_{i,{{j}_{2}}}}$ are completed, we refer to ${{v}_{i+1,{{j}_{0}}}}$ as a child of ${{v}_{i,{{j}_{1}}}}$ and ${{v}_{i,{{j}_{2}}}}$. Tasks ${{v}_{i,{{j}_{1}}}}$ and ${{v}_{i,{{j}_{2}}}}$ are referred to as parents of ${{v}_{i+1,{{j}_{0}}}}$.
\end{defi}

Formally, we define the FFT precedence graph together with its associated notations as follows:
\begin{defi}[FFT Precedence Graph]\label{defi:2}
A precedence graph $G(V,E)$ describes the task dependencies between a set of tasks, where
\begin{enumerate}[1)]
\item Vertex ${{v}_{i,j}}\in V$ represents task $(i,j)$ within $V$.
\item There is an edge $({{v}_{i,{{j}_{1}}}}\to {{v}_{i+1,{{j}_{2}}}})\in E$ if task ${{v}_{i,{{j}_{1}}}}$ must be executed before task ${{v}_{i+1,{{j}_{2}}}}$ can be executed, i.e., ${{v}_{i,{{j}_{1}}}}$ is a parent of ${{v}_{i+1,{{j}_{2}}}}$.
\end{enumerate}
\end{defi}
Since butterfly tasks are executed by processors with the same capacity, the processing times of tasks are the same (i.e., each vertex in the precedence graph takes the same amount of time to execute). We refer to the execution time of a task as one time slot.

With the above backdrop, we now define the compact IFDMA FFT scheduling problem as follows:
\begin{defi}[The FFT Scheduling Problem in Compact IFDMA]\label{defi:3}
Given
\begin{enumerate}[1)]
\item an FFT precedence graph $G(V,E)$ associated with a set of IFDMA streams together with their bin allocations, and
\item $M$ processors, each of which can complete one task in exactly one time slot,
\end{enumerate}
determine a task-execution schedule ${\bf{X}} =\left( {{\chi }_{0}},{{\chi }_{1}},...{{\chi }_{T-1}} \right)$, where ${{\chi }_{t}}$ is the set of tasks to be executed in time slot $t$, and $T$ is the completion time of the collection of tasks, such that,
\begin{enumerate}[1)]
\item the precedence relationships are satisfied;
\item there are no more than $M$ tasks in each of ${{\chi }_{t}}$, and $\underset{t}{\mathop{\bigcup }}\,{{\chi }_{t}}$ consists of all the tasks in the precedence graph;
\item $T$ is minimized.
\end{enumerate}
\end{defi}

Henceforth, unless stated otherwise, we refer to the above FFT scheduling problem in compact IFDMA systems simply as the ``scheduling problem". We refer to an algorithm to solve the scheduling problem as a ``scheduling algorithm".

{\rem{The schedule $\mathbf{X}=\left( {{\chi }_{0}},{{\chi }_{1}},...{{\chi }_{T-1}} \right)$, once determined, may be reused for a duration of time. The reason is as follows. In many applications, the set of IFDMA streams remains the same for a duration of time. As long as the bin allocations do not change, the subcarriers used by each of the streams remain the same. Each stream uses the same subcarriers to transmit its successive IFDMA symbols. An IFDMA FFT needs to be computed for each symbol period for the collection of IFDMA streams. However, the precedence graph remains the same for the successive FFTs. Thus, once we obtain schedule $\mathbf{X}$, the same schedule can be used to perform the IFDMA FFT for the successive IFDMA symbols. In particular, there is no need to compute $\mathbf{X}$ and build new FFT precedence graphs repeatedly. If the set of IFDMA streams changes (e.g., an IFDMA stream ends and leaves, or a new IFDMA stream arrives to join the existing IFDMA stream), a new schedule $\mathbf{X}$ will need to be computed.}}

\subsection {Feasibility of a Task-execution Schedule}
We next describe a step-by-step procedure to check whether a given task-execution schedule $\mathbf{X}=\left\{ {{\chi }_{0}},{{\chi }_{1}},...{{\chi }_{T-1}} \right\}$ is feasible according to criteria 1) and 2) in Definition \ref{defi:3}. Our algorithm (to be described later) follows this step-by-step procedure to construct the schedule to make sure it adheres to the feasibility criteria in each step of the way. In essence, the step-by-step procedure starts with ${{\chi }_{0}}$ and proceed progressively to ${{\chi }_{1}},{{\chi }_{2}},...$ , updating the precedence graph along the way to make sure that the precedence relationships are obeyed each step of the way.

Consider the beginning of time slot $t$, the vertexes in ${{\chi }_{t}}$ must have no parents and there are no more than $M$ tasks in ${{\chi }_{t}}$. If this is not satisfied, then the schedule $\mathbf{X}$ is not feasible. If ${{\chi }_{t}}$ passes the test, we then remove the vertexes in ${{\chi }_{t}}$ together with their output edges from the precedence graph. The updated precedence graph is used to check the feasibility of ${{\chi }_{t+1}}$. We say that $\mathbf{X}$ is a feasible schedule if and only if the precedence graph is nil after vertex removal in time slot $T-1$.

\subsection {Optimality of a Heuristic Scheduling Algorithm}
Given a feasible schedule $\mathbf{X}$, the next question one may ask is whether the schedule is optimal in that the overall computation time $T$ is minimized. Recall from the introduction that the scheduling problems associated with the general non-FFT precedence graph have been proven to be NP-complete \cite{Ref_A1,Ref_A2}. Designing good heuristic algorithms for them with provable bounds is challenging.

The FFT precedence graphs, however, are specific precedence graphs with a regular structure. Section \ref{sec-III} gives a heuristic algorithm that solves the FFT scheduling problem to produce a schedule $\mathbf{X}=\left\{ {{\chi }_{0}},{{\chi }_{1}},...{{\chi }_{T-1}} \right\}$. Section \ref{sec-IV} gives a lower bound on $T$ (denoted by ${{T}^{L}}$) for a given FFT precedence graph. Section \ref{sec-V} shows by experiments that the computation time $T$ of the schedule produced by our heuristic algorithm can reach the lower bound, or is very close to it, with high probability.

In this paper, we characterize the optimality of a heuristic scheduling algorithm by a duple $({{\eta }_{n,M}},{{\gamma }_{n,M}})$ defined as follows.

For a given FFT scheduling problem with ${{2}^{n}}$ subcarriers and $M$ processors, if a scheduling algorithm produces a feasible schedule $\mathbf{X}$ with $T={{T}^{L}}$, then the schedule is optimal. If $T>{{T}^{L}}$, on the other hand, it does not mean the schedule is not optimal, since ${{T}^{L}}$ is just a lower bound. Among all FFT precedence graphs tested, we denote the probability of schedules produced by a heuristic algorithm that satisfies $T={{T}^{L}}$ by ${{\eta }_{n,M}}$. Specifically,
\begin{equation}\label{eq_0A}
\eta_{n,M}=Pr[T=T^L]
\end{equation}
which means the probability of a scheduling algorithm finding the optimal schedule is at least ${{\eta }_{n,M}}$.

For the remaining $1-{{\eta }_{n,M}}$ fraction of the schedules with $T>{{T}^{L}}$, the gap between $T$ and the best possible computation time ${{T}^{*}}$ is $T-{{T}^{*}}\le T-{{T}^{L}}$. The percentage gap between $T$ and ${{T}^{*}}$ is ${\left( T-T^* \right)}/{{{T}^{*}}}\;<{\left( T-{{T}^{L}} \right)}/{{{T}^{L}}}\;$. We define
\begin{equation}\label{eq_0B}
{{\gamma }_{n,M}}=E\left[ {\left( T-{{T}^{L}} \right)}/{{{T}^{L}}}\; \right]
\end{equation}
to be the upper bound of the expected percentage gap, i.e., the average percentage gap between $T$ and ${{T}^{*}}$ is no more than ${{\gamma }_{n,M}}$.

The input to a scheduling algorithm is an FFT precedence graph. As far as scheduling algorithms are concerned, FFT precedence graphs that are isomorphic to each other are equivalent. If a scheduling algorithm finds the optimal schedule for an FFT precedence graph, the optimal scheduling also applies to an isomorphic FFT precedence graph after an isomorphic transformation. Appendix \ref{sec-App2}.1 delves into the isomorphism of FFT precedence graphs.

When running experiments over scheduling algorithms by subjecting them to different FFT precedence graphs, it would be desirable to remove isomorphism so that graphs with many isomorphic instances are not over-represented.

We denote the complete set of FFT precedence graphs by ${{F}_{n}}$, wherein all elements are non-isomorphic. The cardinality of ${{F}_{n}}$ is written as ${{f}_{n}}$.
We refer to the bin allocations that lead to isomorphic FFT precedence graphs as isomorphic bin allocations. Among isomorphic bin allocations, we only select one to put into ${{F}_{n}}$. We refer the reader to Appendix \ref{sec-App2}.1 on how we do so.

Appendix \ref{sec-App2}.2 explains the principle to generate the complete set of non-isomorphic instances ${{F}_{n}}$ for various $n$.
When ${{f}_{n}}$ is not large, it is not an issue to use $F_n$, the complete set, as the test set to conduct experiments and obtain ${{\eta }_{n,M}}$ and ${{\gamma }_{n,M}}$.
However, as shown in Appendix \ref{sec-App2}.2, ${{f}_{n}}$ becomes prohibitively large when $n$ is more than 6.
Table \ref{table:1} lists the $f_n$ values for various $n$, which are derived in Appendix \ref{sec-App2}.2.
\begin{table}[htbp]
\caption{Number of non-isomorphic instances for various $n$}
\begin{center}
\begin{tabularx}{8.0cm}{p{0.1cm}<{\centering}p{1,8cm}<{\centering}p{5.0cm}<{\centering}}
\hline
\noalign{\smallskip}
$n$&FFT Size ($N$)&Number of Non-isomorphic Instances ($f_n$)\\
\noalign{\smallskip}
\hline
\noalign{\smallskip}
\textbf{\textit{$1$}}&$2$&$2$\\
\noalign{\smallskip}
\textbf{\textit{$2$}}&$4$&$4$\\
\noalign{\smallskip}
\textbf{\textit{$3$}}&$8$&$11$\\
\noalign{\smallskip}
\textbf{\textit{$4$}}&$16$&$67$\\
\noalign{\smallskip}
\textbf{\textit{$5$}}&$32$&$2279$\\
\noalign{\smallskip}
\textbf{\textit{$6$}}&$64$&$2598061$\\
\noalign{\smallskip}
\textbf{\textit{$7$}}&$128$&$3.3750E+12$\\
\noalign{\smallskip}
\textbf{\textit{$8$}}&$256$&$5.6952E+24$\\
\noalign{\smallskip}
\textbf{\textit{$9$}}&$512$&$1.6218E+49$\\
\noalign{\smallskip}
\textbf{\textit{$10$}}&$1024$&$1.3151E+98$\\
\noalign{\smallskip}\hline
\end{tabularx}
\label{table:1}
\end{center}
\end{table}

As shown, when $n\ge7$, the number of instances in ${{F}_{n}}$ is so large that it is impractical to generate all the instances and use the complete set of non-isomorphic instances as the test set.
Hence, for $n\ge7$, instead of the full set $F_n$, we randomly generate a subset of $F_n$ (referred to as $F_n^{'}$) to serve as our test set. Appendix \ref{sec-App2}.3 explains how we randomly generate the subset of non-isomorphic instances for large $n$.

With $F_n^{'}$, we then characterize ${{\eta }_{n,M}}$ and ${{\gamma }_{n,M}}$ statistically, as explained in the next few paragraphs.

We first consider how to statistically characterize ${{\eta }_{n,M}}$ for $n\ge7$. For a ${{2}^{n}}$-point IFDMA FFT with $M$ available processors, we randomly sample ${{\xi }_{0}}$ instances with replacement from $F_n$. Define ${{Y}_{{{\xi }_{1}}}}$ to be the event that ${{\xi }_{1}}$ of the ${{\xi }_{0}}$ instances are instances for which the heuristic algorithm can find a solution reaching the lower bound. The conditional probability of ${{Y}_{{{\xi }_{1}}}}$ given ${{\eta }_{n,M}}$ is
\begin{equation}\label{eq_1}
P({{Y}_{{{\xi }_{1}}}}\text{ }\!\!|\!\!\text{ }{{\eta }_{n,M}}\text{)}{=}\left( \begin{matrix}
   {{\xi }_{0}}  \\
   {{\xi }_{1}}  \\
\end{matrix} \right)\eta _{n,M}^{{{\xi }_{1}}}{{(1-{{\eta }_{n,M}})}^{{{\xi }_{0}}-{{\xi }_{1}}}}.
\end{equation}

We are interested in the {\emph{a posteriori}} probability density function (PDF) $p(\left. {{\eta }_{n,M}} \right|{{Y}_{{{\xi }_{1}}}}\text{)}$. We have, by Bayes' rule,
\begin{equation}\label{eq_2}
p({{\eta }_{n,M}}|{{Y}_{{{\xi }_{1}}}}\text{)}=\frac{p({{\eta }_{n,M}})\cdot P({{Y}_{{{\xi }_{1}}}}\text{ }\!\!|\!\!\text{ }{{\eta }_{n,M}}\text{)}}{P({{Y}_{{{\xi }_{1}}}})}.
\end{equation}

However, we do not know the {\emph{a priori}} PDF $p({{\eta }_{n,M}})$. As a conservative measure, we can assume the worst case of having no knowledge on $p({{\eta }_{n,M}})$ and let $p({{\eta }_{n,M}})=1 \text{, } \forall \eta \in [0,1]$. Thus,
\begin{equation}\label{eq_3}
p({{\eta }_{n,M}}|{{Y}_{{{\xi }_{1}}}}\text{)}=\frac{P({{Y}_{{{\xi }_{1}}}}\text{ }\!\!|\!\!\text{ }{{\eta }_{n,M}}\text{)}}{P({{Y}_{{{\xi }_{1}}}})}.
\end{equation}

As a probability density, $p({{\eta }_{n,M}}|{{Y}_{{{\xi }_{1}}}}\text{)}$ must integrate to 1, i.e., $\int{p({{\eta }_{n,M}}|{{Y}_{{{\xi }_{1}}}}\text{)}}d{{\eta }_{n,M}}=1$. This gives,
\begin{equation}\label{eq_4}
\begin{array}{l}
\int {\frac{{P({Y_{{\xi _1}}}{|}{\eta _{n,M}}{{)}}{\kern 1pt} }}{{P({Y_{{\xi _1}}}){\kern 1pt} {\kern 1pt} }}} d{\eta _{n,M}} = 1\\
{\kern 1pt} {\kern 1pt} {\kern 1pt} {\kern 1pt} {\kern 1pt} {\kern 1pt} {\kern 1pt} {\kern 1pt} {\kern 1pt} {\kern 1pt} {\kern 1pt} {\kern 1pt} {\kern 1pt} {\kern 1pt} {\kern 1pt} {\kern 1pt} {\kern 1pt} {\kern 1pt} {\kern 1pt} {\kern 1pt} {\kern 1pt} {\kern 1pt} {\kern 1pt} {\kern 1pt} {\kern 1pt} {\kern 1pt} {\kern 1pt} {\kern 1pt} {\kern 1pt} {\kern 1pt} {\kern 1pt} {\kern 1pt} {\kern 1pt} {\kern 1pt} {\kern 1pt} {\kern 1pt} {\kern 1pt} {\kern 1pt} {\kern 1pt} {\kern 1pt} {\kern 1pt} {\kern 1pt} {\kern 1pt} {\kern 1pt} {\kern 1pt} {\kern 1pt} {\kern 1pt} {\kern 1pt} {\kern 1pt} {\kern 1pt} {\kern 1pt} {\kern 1pt}  \Leftrightarrow P({Y_{{\xi _1}}}) = \int {P({Y_{{\xi _1}}}{{|}}{\eta _{n,M}}{{)}}{\kern 1pt} } d{\eta _{n,M}}
\end{array}.
\end{equation}

Thus, we have
\begin{equation}\label{eq_5}
\begin{aligned}
p({\eta _{n,M}}|{Y_{{\xi _1}}}{{)}}
& {{ = }}\frac{{P({Y_{{\xi _1}}}{{|}}{\eta _{n,M}}{{)}}}}{{\int_0^1 {P({Y_{{\xi _1}}}{{|}}{\eta _{n,M}}{{)}}d{\eta _{n,M}}} }}\\
& = \frac{{\eta _{n,M}^{{\xi _1}}{{(1 - {\eta _{n,M}})}^{{\xi _0} - {\xi _1}}}}}{{\int_0^1 {\eta _{n,M}^{{\xi _1}}{{(1 - {\eta _{n,M}})}^{{\xi _0} - {\xi _1}}}d{\eta _{n,M}}} }}\\
& {{ = }}\frac{{\eta _{n,M}^{{\xi _1}}{{(1 - {\eta _{n,M}})}^{{\xi _0} - {\xi _1}}}}}{{B\left( {{\xi _1} + 1,{\xi _0} - {\xi _1} + 1} \right)}}
\end{aligned}
,\end{equation}
where $B\left( m,n \right)=\int_{0}^{1}{{{x}^{m-1}}{{(1-x)}^{n-1}}dx}$ is the well-known Beta function\footnote{The following mathematical properties of the Beta function, regularized Beta function, and incomplete Beta function can be found in Section 6.2 and Section 26.5 of \cite{BetaHandBook}}.

Let $\alpha$ be the target confidence, let $\eta^L_{n,M}$ be the lower bound of $\eta_{n,M}$, and let $\int_{{{\eta^L_{n,M}}}}^{1}{p({{\eta }_{n,M}}|{{Y}_{{{\xi }_{1}}}}\text{)}} \text{ } d{{\eta }_{n,M}}=\alpha $. We then have
\begin{equation}\label{eq_6}
\begin{array}{l}
\frac{{\int_{{\eta^L_{n,M}}}^1 {\left( \eta _{n,M}\right)^{^{{\xi _1}}}{{(1 - {\eta _{n,M}})}^{{\xi _0} - {\xi _1}}}}\text{ } d{\eta _{n,M}}}}{{B\left( {{\xi _1} + 1,{\xi _0} - {\xi _1} + 1} \right)}} = \alpha \\
\quad\quad\quad\quad
  \Leftrightarrow {I_{{{\eta^L_{n,M}}}}}({\xi _1} + 1,{\xi _0} - {\xi _1} + 1) = 1 - \alpha
\end{array}
,\end{equation}
where
\small
\begin{equation}
{{I}_{{{\eta^L_{n,M}}}}}\left( {{\xi }_{1}}+1,{{\xi }_{0}}-{{\xi }_{1}}+1 \right){=}\frac{B\left( {\eta^L_{n,M}},{{\xi }_{1}}+1,{{\xi }_{0}}-{{\xi }_{1}}+1 \right)}{B\left( {{\xi }_{1}}+1,{{\xi }_{0}}-{{\xi }_{1}}+1 \right)}
\end{equation}
\normalsize
is the regularized Beta function, and
\small
\begin{equation}
B( {\eta^L_{n,M}},{{\xi }_{1}}+1,{{\xi }_{0}}-{{\xi }_{1}}+1 ){=}\int_{0}^{{{\eta^L_{n,M}}}}{{{x}^{{{\xi }_{1}}+1}}{{(1-x)}^{{{\xi }_{0}}-{{\xi }_{1}}+1}}}dx
\end{equation}
\normalsize
is the incomplete beta function. With partial integration, ${{I}_{{\eta^L_{n,M}}}}\left( {{\xi }_1}+{1},{{\xi }_{0}}-{{\xi }_{1}}+1 \right)$ can be further written as,
\begin{equation}\label{eq_7}
\begin{array}{l}
{{I}_{{\eta^L_{n,M}}}}\left( {{\xi }_{1}}+1,{{\xi }_{0}}-{{\xi }_{1}}+1 \right)
\quad \quad \quad \quad \quad
\\ \quad
=\sum\limits_{j={{\xi }_{1}}+1}^{{{\xi }_{0}}+1}{\frac{\left( {{\xi }_{0}}+1 \right)!}{j!\left( {{\xi }_{0}}+1-j \right)!}\left({\eta^L_{n,M}} \right)^{j}{{\left( 1-{{\eta^L_{n,M}}} \right)}^{{{\xi }_{0}}+1-j}}}
\end{array}
.\end{equation}

We then obtain ${{\eta^L_{n,M}}}$ by applying (\ref{eq_7}) to the incomplete Beta function in (\ref{eq_6}).

We next consider how to obtain the $\alpha$-percentile confidence interval of ${{\gamma }_{n,M}}$ for $n\ge7$. For the ${{\xi }_{0}}-{{\xi }_{1}}$ non-lower-bound-reaching schedules (if any), we denote the mean and variance of the observed percentage gap ${\left( T-{{T}^{L}} \right)}/{{{T}^{L}}}\;$ by $\mu $ and ${{\sigma }^{2}}$, respectively. With the central-limit theorem \cite{CentralLimitTheorem}, we construct a random variable $Z$:
\begin{equation}\label{eq_8}
Z=\frac{\mu -{{\gamma }_{n,M}}}{{\sigma }/{\sqrt{{{\xi }_{0}}-{{\xi }_{1}}}}\;}
.\end{equation}
With confidence target $\alpha$ and critical z-score ${{z}_{1-\alpha/2}}$, we have
\begin{equation}\label{eq_9}
P(-{{z}_{1-\alpha /2}}\le Z\le {{z}_{1-\alpha /2}})=\alpha
.\end{equation}

Applying (\ref{eq_8}) to (\ref{eq_9}), we then say,
\begin{equation}\label{eq_10}
\mu -{{z}_{1-\alpha /2}}\cdot \frac{\sigma }{\sqrt{{{\xi }_{0}}-{{\xi }_{1}}}}\le
{{\gamma }_{n,M}}\le \mu +{{z}_{1-\alpha /2}}\cdot \frac{\sigma }{\sqrt{{{\xi }_{0}}-{{\xi }_{1}}}}
\end{equation}
with confidence $\alpha$.

\section{The Multi-Priority Scheduling Algorithm}\label{sec-III}
This section delves into the structural properties of the FFT precedence graph. Leveraging the structural properties, we put forth a heuristic algorithm for the FFT scheduling problem.

\subsection{Structural Properties of the FFT Precedence Graphs}

Several important structural properties of FFT precedence graphs are given below:
\begin{itemize}
\item {\bf{Property 1}} (Number of Parents): Vertexes at stage $0$ have no parent. Vertexes at other stages have two parents located at the preceding stage.
\item {\bf{Property 2}} (Number of Children): Vertexes at stage ${{\log }_{2}}N-1$ (i.e., last stage) have no children. Vertexes at other stages can have one, two, or no children. In Fig. \ref{fig:3}(c), for example, ${{v}_{3,4}}$ at the last stage has no children, while ${{v}_{1,0}}$ has one child, ${{v}_{1,4}}$ has two children, and ${{v}_{2,2}}$ has no children.
\item {\bf{Property 3}} (Parents): Consider vertex ${{v}_{i,j}}$ at stage $i(1\le i\le {{\log }_{2}}N-1)$. Its two parents at stage $i-1$ can be identified as follows:
    \begin{enumerate}[1)]
    \item If $j\bmod \left( {{2}^{{{\log }_{2}}N-i}} \right)<{{2}^{{{\log }_{2}}N-i-1}}$, the two parents are ${{v}_{i{-}1,j}}$ and ${{v}_{i{-}1,j+{{2}^{{{\log }_{2}}N-i-1}}}}$.
    \item If $j\bmod \left( {{2}^{{{\log }_{2}}N-i}} \right)\ge {{2}^{{{\log }_{2}}N-i-1}}$, the two parents are ${{v}_{i{-}1,j-{{2}^{{{\log }_{2}}N-i-1}}}}$ and ${{v}_{i{-}1,j}}$.
    \end{enumerate}
\item  {\bf{Property 4}} (Children): Consider ${{v}_{i,j}}$ at stage $i(0\le i\le {{\log }_{2}}N-2)$ . Its children at stage $i+1$ (if any) can be identified as follows:
    \begin{enumerate}[1)]
    \item If $j\bmod \left( {{2}^{{{\log }_{2}}N-i-1}} \right)<{{2}^{{{\log }_{2}}N-i-2}}$, the potential children are ${{v}_{i{+}1,j}}$ and ${{v}_{i{+}1,j+{{2}^{{{\log }_{2}}N-i-2}}}}$.
    \item If $j\bmod \left( {{2}^{{{\log }_{2}}N-i-1}} \right)\ge {{2}^{{{\log }_{2}}N-i-2}}$, the potential children are ${{v}_{i{+}1,j-{{2}^{{{\log }_{2}}N-i-2}}}}$ and ${{v}_{i{+}1,j}}$.
    \end{enumerate}
\end{itemize}

Note that one or both of the potential children may be absent in the FFT precedence graph.
The above properties lead to an observation that the FFT butterfly structure in turn induces a certain partial butterfly pattern in the precedence graph.
This is, if ${{v}_{i,{{j}_{1}}}}$ has two children, there must be another vertex ${{v}_{i,{{j}_{2}}}}$ sharing the same two children vertexes with ${{v}_{i,{{j}_{1}}}}$ (proved as Corollary 1 in Appendix \ref{sec-App4}). If ${{v}_{i,{{j}_{1}}}}$ has only one child, obviously, there is another parent at stage $i$ sharing the same child with ${{v}_{i,{{j}_{1}}}}$.

With the above observations, we give the definitions of ``vertex pair" and ``the companion of a vertex" below.
\begin{defi}[Vertex Pair and the Companion of a Vertex]\label{defi:4}
If a parent vertex shares children with another parent vertex, we refer to the two parent vertexes as a vertex pair. Each of the parent vertexes is the companion of the other.
\end{defi}

\begin{figure}[htbp]
  \centering
  \includegraphics[width=3.25in]{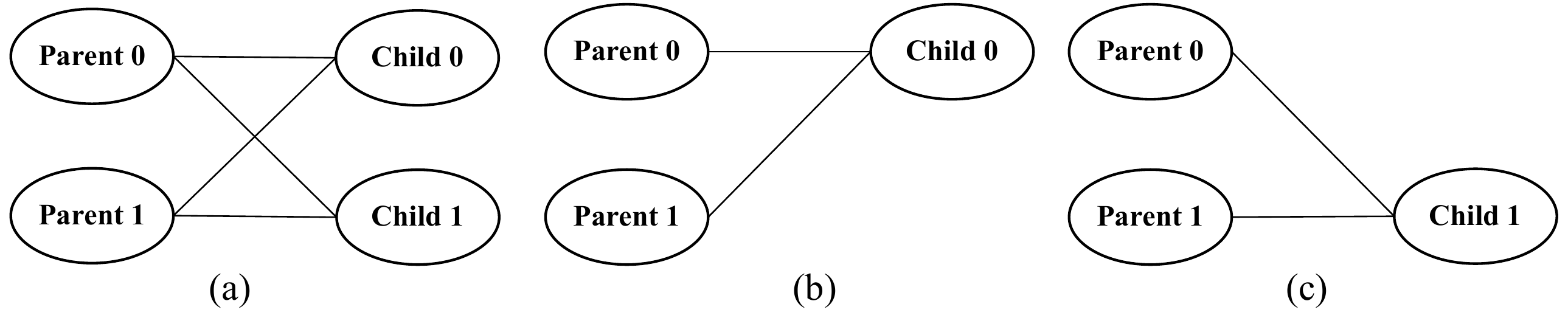}\\
  \caption{Legitimate parent-child structures within the FFT precedence graph.}
\label{fig:5}
\end{figure}

Fig. \ref{fig:5} illustrates the legitimate parent-child structures within the FFT precedence graph. A vertex can have zero, one or two children. Fig. \ref{fig:5}(a) shows the case of two vertexes sharing two common children. Fig. \ref{fig:5}(b) and (c) show the cases of two vertexes, each having one child, and that they share this common child.

Note that the structures in Fig. \ref{fig:5} are only applicable to the initial FFT precedence graph before the scheduling algorithm is executed. This is because if a parent vertex is selected for execution and then removed from the FFT precedence graph, the structure of the graph changes. Vertex pairs and the companion relationship in Definition \ref{defi:4}, on the other hand, are applicable to the precedence graph throughout the execution of the scheduling algorithm. If vertex A is the companion of vertex B, and it is selected for execution before vertex B, we refer to vertex A as a selected companion of vertex B.

\subsection{The Heuristic Scheduling Algorithm}
\begin{figure*}[htbp]
  \centering
  \includegraphics[width=7in]{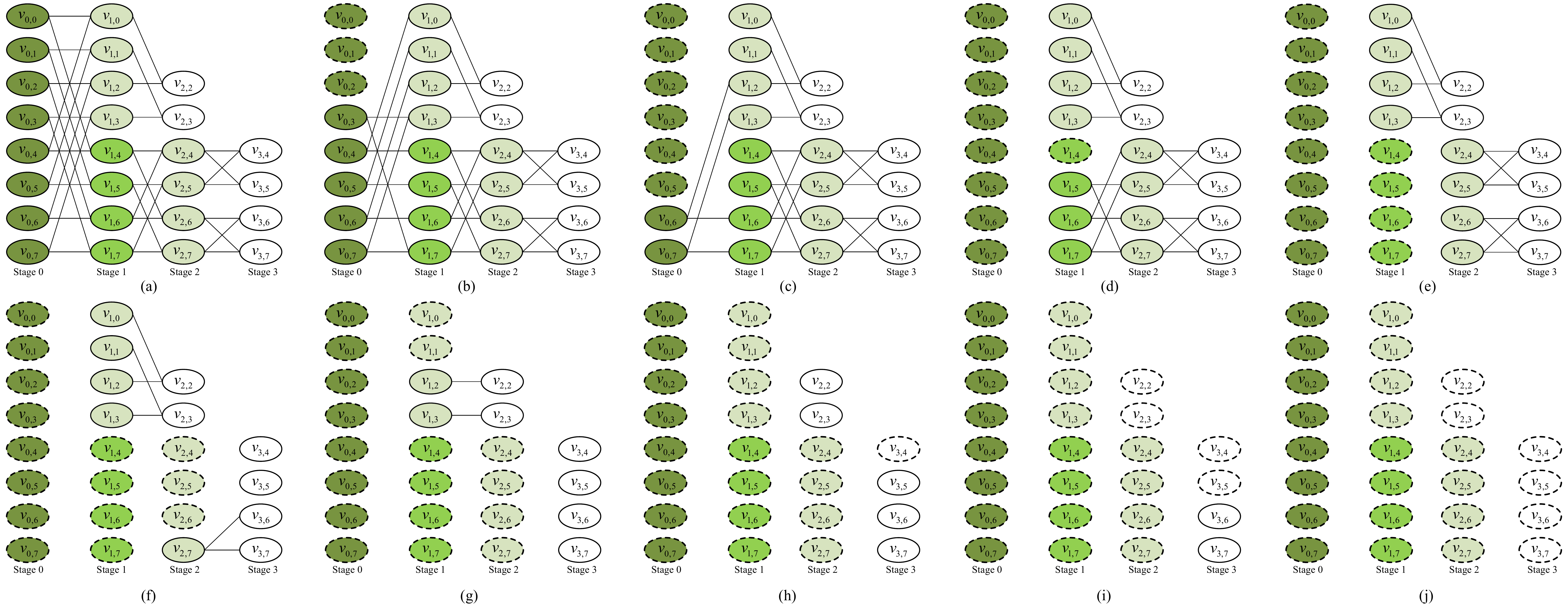}\\
  \caption{Evolution of the FFT precedence graph during the execution of the heuristic algorithm, assuming three processors. The vertexes enclosed by a solid boundary belong to the precedence graph. The dashed vertexes do not belong to the precedence graph, and they are shown here for illustrative purposes only.}
\label{fig:6}
\end{figure*}

We now construct a heuristic scheduling algorithm that exploits the structure of the FFT precedence graph. Fig. \ref{fig:6} presents an example of the evolution of the FFT precedence graph during the execution of our heuristic algorithm. The example assumes the same IFDMA bin allocation as in Fig. \ref{fig:3}(a) of Section \ref{sec-II}. The vertexes are marked in different colors to facilitate later explanation of our heuristic algorithm. Here, we assume that the number of processors is three. A scheduling algorithm begins with the initial precedence graph shown in Fig. \ref{fig:6}(a) and ends with the final precedence graph  shown in Fig. \ref{fig:6}(j). Each of the subfigures from Fig. \ref{fig:6}(b) to Fig. \ref{fig:6}(i) shows an intermediate FFT precedence graph at the end of a time slot. When a vertex is removed from the FFT precedence graph, we mark it as a dashed ellipse, the removed edges are simply not shown.

We refer to our scheduling algorithm as multi-priority scheduling (MPS). As the name suggests, MPS associates each vertex (task) with a priority vector $\left\langle {{P}_{1}},{{P}_{2}},...,{{P}_{H}} \right\rangle$ to determine its execution priority. In particular, the priority element ${{P}_{i}}$ takes precedence over the priority element ${{P}_{i+1}}$ when comparing the priorities of two vertexes. For example, if the ${{P}_{1}},{{P}_{2}},...,{{P}_{i-1}}$ of tasks A and task B are the same, but ${{P}_{i}}$ of task A is higher than of task B, then task A is of higher priority than task B.

Throughout this paper, we study the MPS algorithm for which the priority vector $\left\langle {{P}_{1}},{{P}_{2}},{{P}_{3}},{{P}_{4}} \right\rangle$ has four elements. This algorithm was applied in the example of Fig. \ref{fig:6}. Let us now specify the elements in $\left\langle {{P}_{1}},{{P}_{2}},{{P}_{3}},{{P}_{4}} \right\rangle$ and explain how they are used to yield the scheduling results in Fig. \ref{fig:6}.

First, ${{P}_{1}}$ of a vertex is the number of generations of descendants it has, wherein ``descendants" refers to its children, its children's children, and so on. Fig. \ref{fig:6}(a) illustrates how we count the number of generations. We mark the vertexes in Fig. \ref{fig:6}(a) with four different colors according to how many generations of descendants they have. Vertexes with no child are colored in white. Their parents have one generation of descendants, and they are colored in light green. Vertexes in green have two generations of descendants, and vertexes in dark green have three generations of descendants.

The intuition for setting ${{P}_{1}}$ as above is as follows. Recall that the children of a vertex must be executed in a different time slot than the vertex. Thus, the number of generations of descendants of a vertex corresponds to the number of extra time slots that are needed in addition to the time slot used to execute the vertex. It will be advantageous to select the vertex with more generations of descendants to free up the dependencies of its descendants on it.

Let us look at Fig. \ref{fig:6}. The precedence graph after time slot 2 is shown in Fig. \ref{fig:6}(c). Three tasks can be selected for execution in time slot 3. Among the eight executable vertexes (${{v}_{0,6}},{{v}_{0,7}},{{v}_{1,0}},...{{v}_{1,5}}$) in Fig. \ref{fig:6}(c), MPS selects all of the vertexes with high ${{P}_{1}}$ value (${{P}_{1}}{(}{{v}_{0,6}}{)=}{{P}_{1}}{(}{{v}_{0,7}}{)=}3$) and one of the vertexes with medium ${{P}_{1}}$ value (${{P}_{1}}{(}{{v}_{1,4}}{)=2}$). Our algorithm does not select the light green vertexes in time slot 3 because their ${{P}_{1}}$ is lower than those in green and dark green.

Next, we present the specification of ${{P}_{2}}$. We divide the vertexes into three types. Childless type refers to vertexes with no children. Paired type refers to vertexes for which both the vertex itself and its companion have not yet been selected for execution (see Definition \ref{defi:4} for the concept of companion). Singleton type refers to unselected vertexes whose companion has already been selected for execution in a prior time slot. Accordingly, ${{P}_{2}}$ is set to 0, 1, and 2, respectively.

The intuition for setting ${{P}_{2}}$ as above is as follows. Since the selection of childless vertexes brings no new executable vertex in the next time slot, we give it the lowest ${{P}_{2}}$. For paired vertexes, it takes two processors to execute them together in the same time slot and obtain new executable vertex(es) (i.e., the children) in the next slot. For singleton vertexes, it takes only one processor to obtain the same number of new executable vertex(es) in the next slot. So, we give the middle and the highest ${{P}_{2}}$ to paired vertex and singleton vertex, respectively. Note that ${{P}_{2}}$ is a myopic measure while ${{P}_{1}}$ looks further ahead. Hence, ${{P}_{1}}$ takes precedence over ${{P}_{2}}$.

Let us look at the example of Fig. \ref{fig:6} again. The precedence graph at the end of time slot 3 is shown in Fig. \ref{fig:6}(d). Both ${{v}_{1,5}}$, ${{v}_{1,6}}$ and ${{v}_{1,7}}$ have the highest ${{P}_{1}}$ among the unselected vertexes. Note that ${{v}_{1,5}}$ and ${{v}_{1,7}}$ are of paired type, because they form a pair and neither of them has been selected for execution, while ${{v}_{1,6}}$ is of singleton type because its companion ${{v}_{1,4}}$ has been selected for execution in a previous time slot. Hence, the selection order of the following time slot (time slot 4) is ${{v}_{1,6}}\to {{v}_{1,5}}\to {{v}_{1,7}}$. If $M$ were to be 2, then only ${{v}_{1,6}}$ and one of ${{v}_{1,5}}$ or ${{v}_{1,7}}$ would be selected; however, since $M=3$ in our example, all of them are selected.

Next, ${{P}_{3}}$ of a vertex is the number of children it has. Note that both a singleton-type vertex and a pair of paired-type vertexes may have two children in the next time slot. But an intelligent algorithm should select the former for execution first. Hence, ${{P}_{2}}$ takes precedence over ${{P}_{3}}$.

In Fig. \ref{fig:6}, the precedence graph at the end of time slot 5 is shown in Fig. \ref{fig:6}(e). All vertexes in light green are executable, and they have the same ${{P}_{1}}$ and ${{P}_{2}}$ values. Our MPS algorithm prefers ${{v}_{2,4}}$,${{v}_{2,5}}$, ${{v}_{2,6}}$ and ${{v}_{2,7}}$, because each of them has two children, while each of the other four executable vertexes (${{v}_{1,0}}$,${{v}_{1,1}}$, ${{v}_{1,2}}$ and ${{v}_{1,3}}$) has only one child. Since there are only three processors, the algorithm selects three vertexes among  ${{v}_{2,4}}$,${{v}_{2,5}}$, ${{v}_{2,6}}$ and ${{v}_{2,7}}$. Hence, the selection order of the following time slot (time slot 6) is ${{v}_{2,4}}\to {{v}_{2,5}}\to {{v}_{2,6}}$.

Next, if executable vertexes have the same ${{P}_{1}}$, ${{P}_{2}}$ and ${{P}_{3}}$ values, we can randomly select the vertexes to execute. However, we are interested in a deterministic algorithm (for analytical purposes to establish a performance bound, as will be elaborated later). Thus, we have ${{P}_{4}}$ to break the tie. We select those vertexes from the top of the precedence graph to the bottom of the precedence graph. So, we assign ${{P}_{4}}$ to a vertex according to the row of the precedence graph at which it is located. Specifically, ${{P}_{4}}=({N}/{2}\;-1)-j$, where $j$ is the row that ${{v}_{i,j}}$ lies in.

In Fig. \ref{fig:6}, the precedence graph at the end of time slot 7 is shown in Fig. \ref{fig:6}(h). There are five executable vertexes with the same ${{P}_{1}}$, ${{P}_{2}}$ and ${{P}_{3}}$ values. Among them, ${{v}_{2,2}}$ is located at the highest row, with its ${{P}_{4}}$ equal to 5. Hence, the selection order of the next time slot (time slot 8) is ${{v}_{2,2}}\to {{v}_{2,3}}\to {{v}_{3,5}}$.

With the above setup, when comparing the priority of two vertexes, we are actually comparing their priority vectors. For realization, it will be convenient to compare two scalar values rather than two vectors. In other words, we want to map a priority vector $\left\langle {{P}_{1}(v_{i,j})},{{P}_{2}(v_{i,j})},{{P}_{3}(v_{i,j})},{{P}_{4}(v_{i,j})} \right\rangle$ to a priority scalar $P(v_{i,j})$ in such a way that the priority order is preserved. As shown in Appendix \ref{sec-App5}, a possible mapping scheme from a priority vector to a priority scalar is,
\begin{equation}\label{eq_16}
P({{v}_{i,j}})=\sum\limits_{h=1}^{H=4}{{{\left( \frac{N}{2} \right)}^{H-h}}{{P}_{h}}({{v}_{i,j}})}
.\end{equation}

Note that for our $H=4$ MPS, the values of ${{P}_{1}},{{P}_{3}},{{P}_{4}}$ are fixed at the beginning of the scheduling and will never change as the algorithm proceeds. However, ${{P}_{2}}$ of a vertex ${{v}_{i,j}}$ may change dynamically, according to whether the companion vertex of ${{v}_{i,j}}$ has already been selected for execution.

Recall that a paired-type vertex shares one or two common children with another companion vertex. If a vertex has children in the initial precedence graph, it has an unselected companion vertex. Hence, there are only two types of vertexes in the initial FFT precedence graph: the childless type and the paired type. With the execution of our MPS algorithm, a paired vertex ${{v}_{i,j}}$ will change to a singleton vertex once its companion has been selected for execution. As a result, ${{P}_{2}}$ of a vertex can be 0 or 1 in the initial FFT precedence graph (depends on whether it has children). Then, with the execution of MPS, the ${{P}_{2}}$ of a vertex can be adjusted to 2 accordingly.

Formally, the above description of the assignment of priority vectors, and the serialization for the computation of the priority scalar, can be summarized as follows:
For ${{v}_{i,j}}\in V$, we denote the set of ${{v}_{i,j}}$'s children by $K({{v}_{i,j}})$ and the cardinality of $K({{v}_{i,j}})$ by ${{k}_{i,j}}$, we then have
\begin{enumerate}[1)]
    \item ${{P}_{1}}$ assignment: if ${{k}_{i,j}}=0$, then ${{P}_{1}}({{v}_{i,j}})=0$; else ${{P}_{1}}({{v}_{i,j}})=\underset{{{v}_{i+1,x}}\in K({{v}_{i,j}})}{\mathop{\max }}\,{{P}_{1}}({{v}_{i+1,x}}) +1$.
    \item ${{P}_{2}}$ assignment (initial value): For ${{v}_{i,j}}$, if ${{k}_{i,j}}=0$, then ${{P}_{2}}({{v}_{i,j}})=0$; else  ${{P}_{2}}({{v}_{i,j}})=1$.
    \item ${{P}_{2}}$ assignment (dynamically adjusted value): If ${{v}_{i,j}}$'s companion is newly selected for execution, let ${{P}_{2}}({{v}_{i,j}}) = 2$.
    \item ${{P}_{3}}$ assignment: For ${{v}_{i,j}}$,  ${{P}_{3}}({{v}_{i,j}})={{k}_{i,j}}$.
    \item ${{P}_{4}}$ assignment: For ${{v}_{i,j}}$,  ${{P}_{4}}({{v}_{i,j}})=\frac{N}{2}-1-j$.
    \item Priority scalar calculation: $P({{v}_{i,j}}){=}\sum\limits_{h=1}^{H\text{=}4}{{{\left( \frac{N}{2} \right)}^{H-h}}{{P}_{h}}({{v}_{i,j}})}$.
\end{enumerate}

In Appendix \ref{sec-App3}, we describe the $H=4$ MPS algorithm with pseudocode.

\section{IFDMA FFT Computation-time Lower Bound}\label{sec-IV}
This section derives a general lower bound ${{T}^{L}}$ on the IFDMA FFT computation time. Any feasible schedule ${\bf{X}} =\left( {{\chi }_{0}},{{\chi }_{1}},...{{\chi }_{T-1}} \right)$ constrained by the FFT precedence graph and the number of available processors must have computation time $T\ge {{T}^{L}}$.

A trivial lower bound is,
\begin{equation}\label{eq_17}
{T^L} = \left\lceil {{{\sum\limits_{i = 0}^{\log N - 1} {{Q_i}} } \mathord{\left/
 {\vphantom {{\sum\limits_{i = 0}^{\log N - 1} {{Q_i}} } M}} \right.
 \kern-\nulldelimiterspace} M}} \right\rceil
,\end{equation}
where $\left\lceil \cdot  \right\rceil $ is the round-up operation, and ${{Q}_{i}}$ is the number of vertexes at stage $i$. This is, however, a rather loose lower bound.

In the following, we prove a tight lower bound for $T$. Our argument divides the vertexes into ``trunk vertexes" and ``branch vertexes", defined as follows:
\begin{defi}[Trunk Vertex and Branch Vertex]\label{defi:5}
In an initial FFT precedence graph with $\beta $ FFT stages\footnote{Since many bin allocations do not require tasks in later FFT stages, we have $\beta \le {{\log }_{2}}N$ in general.}, if a vertex ${{v}_{i,j}}$ satisfies,
\begin{equation}\label{eq_18}
{{P}_{1}}({{v}_{i,j}})=\beta -1-i
,\end{equation}
we refer to it as a trunk vertex; otherwise, we refer to it as a branch vertex.
\end{defi}

Recall that in Section \ref{sec-III}, we define ${{P}_{1}}({{v}_{i,j}})$ to be the number of generations of descendants vertex ${{v}_{i,j}}$ has. Since vertex ${{v}_{i,j}}$ is located at stage $i$, and the maximum number of remaining stages after stage $i$ is $\beta -1-i$, in general we have ${{P}_{1}}({{v}_{i,j}})\le \beta -1-i$. Hence, trunk vertexes at stage $i$ are vertexes at stage $i$ that have the maximum possible ${{P}_{1}}$.

\begin{figure}[htbp]
  \centering
  \includegraphics[width=2.0in]{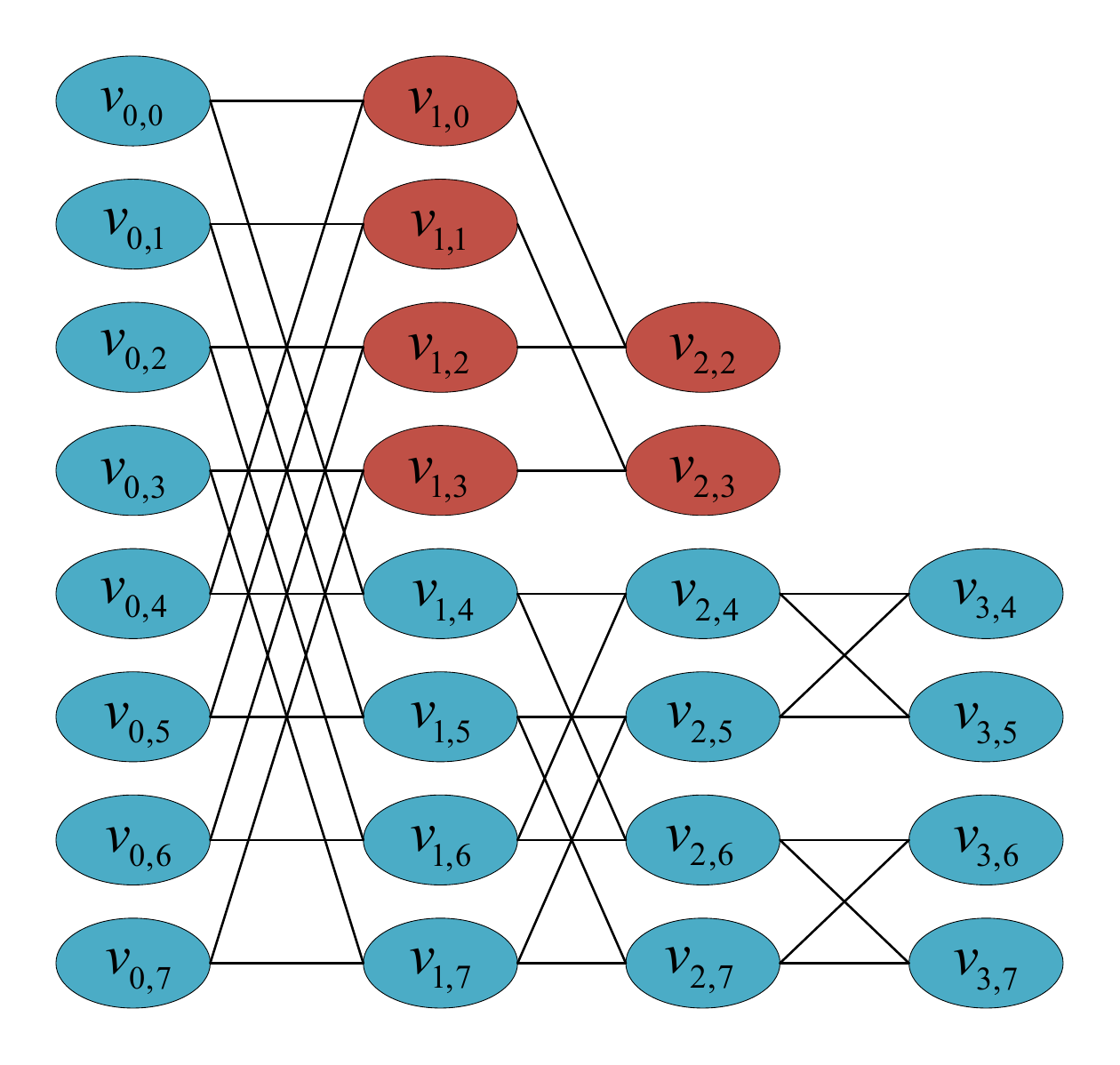}\\
  \caption{An example of ``trunk vertexes" and ``branch vertexes".}
\label{fig:7}
\end{figure}

Fig. \ref{fig:7} gives an illustration. Vertexes in blue are the trunk vertexes, and vertexes in red are the branch vertexes. The rationale for our nomenclature is as follows. We picture vertexes in blue to be part of the main trunk of a tree and vertexes in red to be part of branches of a tree.

\begin{lem}\label{lamma:1}
A trunk vertex ${{v}_{i,j}}$ with $i\in \{0,1,...\beta -2\}$, that is not located at the last stage (i.e., stage $\beta -1$) has at least one child that is also a trunk vertex.
\end{lem}
\begin{NewProof}
Since ${{v}_{i,j}}$ is a trunk vertex, we have
\begin{equation}\label{eq_19}
{{P}_{1}}({{v}_{i,j}})=1+\underset{{{v}_{i+1,x}}\in K({{v}_{i,j}})}{\mathop{\max }}\,{{P}_{1}}({{v}_{i+1,x}})=\beta -i-1
,\end{equation}
where $K({{v}_{i,j}})$ is the children set of ${{v}_{i,j}}$.

From (\ref{eq_19}), we know that at least one vertex ${{v}_{i+1,x}}$ satisfies,
\begin{equation}\label{eq_20}
{{\left. {{P}_{1}}({{v}_{i+1,x}}) \right|}_{{{v}_{i+1,x}}\in K({{v}_{i,j}})}}=\beta -(i+1)-1
,\end{equation}
i.e., ${{v}_{i+1,x}}$ is a trunk vertex.
\end{NewProof}

Let ${{T}_{tr}}(i)$, $i=0,1,...\beta -2$, be the time slot by which all trunk vertexes at stage $i$ are executed.

\begin{lem}\label{lamma:2}
${{T}_{tr}}(i+1)\ge {{T}_{tr}}(i)+1$.
\end{lem}
\begin{NewProof}
In general, some trunk vertexes at stage $i$ could be executed at time slots earlier than ${{T}_{tr}}(i)$. But there must be at least one trunk vertex ${{v}_{i,j}}$ at stage $i$ that is executed in time slot ${{T}_{tr}}(i)$. From Lemma \ref{lamma:1}, trunk vertex ${{v}_{i,j}}$ must have one child trunk vertex, say vertex ${{v}_{i+1,x}}$, at stage $i+1$ that depends on ${{v}_{i,j}}$. The earliest time slot by which ${{v}_{i+1,x}}$ can be executed is ${{T}_{tr}}(i)+1$. Thus, in general, ${{T}_{tr}}(i+1)\ge {{T}_{tr}}(i)+1$.
\end{NewProof}

Let ${{R}_{i}}$ be the number of trunk vertexes at stage $i$. Some stages have more than $M$ trunk vertexes, and some stages have no more than $M$ trunk vertexes. Let $\mathbf{U}$ be the set of stages with more than $M$ trunk vertexes, and $\mathbf{W}$ be the set of stages with no more than $M$ trunk vertexes.

\begin{lem}\label{lamma:3}
In an FFT precedence graph, the stages in $\mathbf{U}$ precedes the stages in $\mathbf{W}$.
\end{lem}
\begin{NewProof}
A pair of trunk parent vertexes at stage $i$ have no more than two child trunk vertexes at stage $i+1$. Each child trunk vertexes, on the other hand, must have two parent trunk vertexes.
Therefore, ${{R}_{i}}$, the number trunk vertexes at stage $i$, can be no smaller than ${{R}_{i+1}}$, the number of trunk vertexes at stage $i+1$. So, the stages in $\mathbf{U}$ precede the stages in $\mathbf{W}$.
\end{NewProof}

We denote the cardinality of the set $\mathbf{W}$ by $m$ in the following. We also refer to the computation time required to execute all trunk vertexes as ${{T}_{tr}}$, and we denote the lower bound of ${{T}_{tr}}$ by $T_{tr}^{L}$. With the above lemmas, we obtain $T_{tr}^{L}$ as follows.

\begin{lem}\label{lamma:4}
${{T}_{tr}}\ge T_{tr}^{L}\triangleq \left\lceil {{{\sum\limits_{i = 0}^{\beta  - m - 1} {{R_i}} } \mathord{\left/
 {\vphantom {{\sum\limits_{i = 0}^{\beta {-}m{-}1} {{R_i}} } M}} \right.
 \kern-\nulldelimiterspace} M}} \right\rceil  + m$
.\end{lem}
\begin{NewProof}
From Lemma \ref{lamma:3}, the first $\beta -m$ stages have more than $M$ in each stage, and their execution can finish no earlier than,
\begin{equation}\label{eq_21}
{{T}_{tr}}\left( \beta -m-1 \right)\ge  \left\lceil {{{\sum\limits_{i = 0}^{\beta  - m - 1} {{R_i}} } \mathord{\left/
 {\vphantom {{\sum\limits_{i = 0}^{\beta {-}m{-}1} {{R_i}} } M}} \right.
 \kern-\nulldelimiterspace} M}} \right\rceil
.\end{equation}
Applying Lemma \ref{lamma:2} to stage $\beta -m$ , we have
\begin{equation}\label{eq_22}
{{T}_{tr}}\left( \beta {-}m \right)\ge {{T}_{tr}}\left( \beta {-}m{-}1 \right){+}1\ge  \left\lceil {{{\sum\limits_{i = 0}^{\beta  - m - 1} {{R_i}} } \mathord{\left/
 {\vphantom {{\sum\limits_{i = 0}^{\beta {-}m{-}1} {{R_i}} } M}} \right.
 \kern-\nulldelimiterspace} M}} \right\rceil {+}1
.\end{equation}
Similarly, for stage $\beta -m+1$, we have ${{T}_{tr}}\left( \beta -m+1 \right)\ge {{T}_{tr}}\left( \beta -m \right){+}1$. And so on and so forth. The last trunk vertexes at the stage $\beta -1$ can therefore finish execution no earlier than $\left\lceil {{{\sum\limits_{i = 0}^{\beta  - m - 1} {{R_i}} } \mathord{\left/
 {\vphantom {{\sum\limits_{i = 0}^{\beta {-}m{-}1} {{R_i}} } M}} \right.
 \kern-\nulldelimiterspace} M}} \right\rceil +m$. Thus, we have
\begin{equation}\label{eq_23}
{T_{tr}} \ge T_{tr}^L \buildrel \Delta \over = \left\lceil {{{\sum\limits_{i = 0}^{\beta  - m - 1} {{R_i}} } \mathord{\left/
 {\vphantom {{\sum\limits_{i = 0}^{\beta {-}m{-}1} {{R_i}} } M}} \right.
 \kern-\nulldelimiterspace} M}} \right\rceil  + m
.\end{equation}
\end{NewProof}

The $T_{tr}^{L}$ in (\ref{eq_23}) would be a lower bound for the overall problem if there were no branch vertexes. We next take into consideration the computation time needed by the branch vertexes. When packing the vertexes into the schedule implicit in our procedure for obtaining $T_{tr}^{L}$ above, there could be time slots in which there are fewer than $M$ vertexes being executed. As far as a lower bound for the overall computation time is concerned, we could imagine that we could pack branch vertexes into such time slots for execution without constraints of the precedence graph. In other words, we imagine that the branch vertexes could fully exploit the unused capacities of the time slots allocated to the trunk vertexes. The lower bound obtained as such may not be achievable, but it is still valid as far as a lower bound.

The total number of branch vertexes is $\sum\limits_{i=0}^{\beta -1}{\left( {{Q}_{i}}-{{R}_{i}} \right)}$. If the unused capacities is not enough for executing all branch vertexes, i.e., $M\cdot T_{tr}^{L}-\sum\limits_{i=0}^{\beta -1}{{{R}_{i}}}<\sum\limits_{i=0}^{\beta -1}{\left( {{Q}_{i}}-{{R}_{i}} \right)}$, we need extra time slots after $T_{tr}^{L}$ to execute the remaining branch vertexes. We refer to the computation time for executing the remaining branch vertexes as ${{T}_{br}}$, and we denote the lower bound of ${{T}_{br}}$ by $T_{br}^{L}$.

\begin{lem}\label{lamma:5}
${{T}_{br}}\ge T_{br}^{L}\triangleq \left\lceil {l}/{M} \right\rceil$, where
$l=\max \left\{{0, \sum\limits_{i = 0}^{\beta  - 1} {{Q_i}}  - M\left\lceil {{{\sum\limits_{i = 0}^{\beta  - m - 1} {{R_i}} } \mathord{\left/
 {\vphantom {{\sum\limits_{i = 0}^{\beta  - m - 1} {{R_i}} } M}} \right.
 \kern-\nulldelimiterspace} M}} \right\rceil  + Mm} \right\}.$
\end{lem}
\begin{NewProof}
There are $M\cdot T_{tr}^{L}$ processor capacities in the first $T_{tr}^{L}$ time slots. During the $T_{tr}^{L}$ time slots, the unused capacities are given by,
\begin{equation}\label{eq_24}
E=M\cdot T_{tr}^{L}-\sum\limits_{i=0}^{\beta -1}{{{R}_{i}}}
.\end{equation}
After packing branch vertexes into $T_{tr}^{L}$ time slots to used up the capacities $E$, the number of remaining branch vertexes not packed into the $T_{tr}^{L}$ time slots is,
\begin{equation}\label{eq_25}
\begin{aligned}
l&=\max \left\{ 0,\sum\limits_{i=0}^{\beta -1}{\left( {{Q}_{i}}-{{R}_{i}} \right)}-E \right\} \\
 & {=}\max \left\{ 0,\sum\limits_{i=0}^{\beta -1}{{{Q}_{i}}}-M\cdot T_{tr}^{L} \right\} \\
 & {=}\max \left\{ 0,\sum\limits_{i=0}^{\beta -1}{{{Q}_{i}}}-M\left\lceil {\sum\limits_{i=0}^{\beta -m-1}{{{R}_{i}}}}/{M}\; \right\rceil +Mm \right\} \\
\end{aligned}
.\end{equation}

Thus, ${{T}_{br}}$ is lower bounded by,
\begin{equation}\label{eq_26}
{{T}_{br}}\ge T_{br}^{L}\triangleq \left\lceil \frac{l}{M} \right\rceil
.\end{equation}
\end{NewProof}

\begin{thm}\label{theorem:1}
${{T}}\ge T^{L} \triangleq  \left\lceil {{{\sum\limits_{i = 0}^{\beta  - m - 1} {{R_i}} } \mathord{\left/
 {\vphantom {{\sum\limits_{i = 0}^{\beta  - m - 1} {{R_i}} } M}} \right.
 \kern-\nulldelimiterspace} M}} \right\rceil + \left\lceil {l}/{M} \right\rceil + m .$
\end{thm}
\begin{NewProof}
Since $T \ge T_{tr}^{L}+T_{br}^{L}$, we prove Theorem \ref{theorem:1} with Lemma \ref{lamma:4} and Lemma \ref{lamma:5}.
\end{NewProof}

\section{Experiments and Analysis}\label{sec-V}

\setcounter{table}{1}

We refer to the IFDMA FFT computed according to the schedule found by MPS as ``MPS-FFT". The first subsection presents experimental results showing that the schedule found by the MPS algorithm is close-to-optimal in terms of the FFT computation time. The second subsection presents experimental results showing that MPS-FFT has a shorter computation time than conventional FFT.

\subsection {Optimality of MPS}
This subsection focuses on the optimality of the schedules obtained by MPS. Recall from Section \ref{sec-II} that we use ${{\eta }_{n,M}}$ to denote the probability of finding a schedule whose computation time reaches the lower bound in an FFT scheduling problem with ${{2}^{n}}$ subcarriers and $M$ processors. We assume a fully packed IFDMA system, in which all subcarriers are used. If the computation time does not reach the lower bound, we then use ${{\gamma }_{n,M}}$ to denote the average percentage gap between the observed computation time and the theoretical lower bound.

{\noindent{\bf{Case 1: arbitrary number of processors}}}

Let us first assume an arbitrary number of processors between one and ${{2}^{n-1}}$, the maximum processors needed.
We denote the average value of ${{\eta }_{n,M}}$ and ${{\gamma }_{n,M}}$ over $M$ by ${{\eta }_{n}}$ and ${{\gamma }_{n}}$, respectively.
The experimental results of ${{\eta }_{n}}$ and ${{\gamma }_{n}}$ are given in Table II.
As explained in Section \ref{sec-II}, when the number of subcarriers is large, the number of non-isomorphic instances for IFDMA can be huge.
Hence, when $n\ge7$, we randomly sampled ${{\xi }_{0}}={f_6}=2598061$ of ${{f}_{n}}$ non-isomorphic instances for testing, where ${{f}_{n}}$ is the total number of non-isomorphic instances when there are ${{2}^{n}}$ subcarriers.
The method to generate non-isomorphic instances with equal probability is presented in Appendix \ref{sec-App2}.3.
We follow the analysis in Section \ref{sec-II} to statistically obtain ${{\eta }_{n,M}}$ and ${{\gamma }_{n}}$ with a confidence level of $\alpha =0.95$.
For systems of 8 to 64 subcarriers, we tested all non-isomorphic instances. The two-point and four-point FFT modules are trivial and are omitted in the experiments.
\begin{table}[htbp]\label{table:2}
\caption{${{\eta }_{n}}$ and ${{\gamma }_{n}}$ for the scheduling problems in a ${{2}^{n}}$-point IFDMA FFT (the case of an arbitrary number of processors)}
\begin{center}
\begin{tabularx}{6.0cm}{p{0.5cm}<{\centering}p{3.1cm}<{\centering}p{1.4cm}<{\centering}}
\hline
$n$ & ${\eta _n}$ & ${\gamma _n}$ \\
\hline
\noalign{\smallskip}
\textbf{\textit{$3$}}&$1.0000$&$0.0000$\\
\noalign{\smallskip}
\textbf{\textit{$4$}}&$0.9985$&$0.0250$\\
\noalign{\smallskip}
\textbf{\textit{$5$}}&$0.9984$&$0.0460$\\
\noalign{\smallskip}
\textbf{\textit{$6$}}&$0.9981$&$0.0648$\\
\noalign{\smallskip}
\textbf{\textit{$7$}}&$\ge 0.9972$&$\le 0.0391$\\
\noalign{\smallskip}
\textbf{\textit{$8$}}&$\ge 0.9942$&$\le 0.0146$\\
\noalign{\smallskip}
\textbf{\textit{$9$}}&$\ge 0.9901$&$\le 0.0121$\\
\noalign{\smallskip}
\textbf{\textit{$10$}}&$\ge 0.9870$&$\le 0.0104$\\
\noalign{\smallskip}\hline
\end{tabularx}
\end{center}
\end{table}

We can see from Table II that MPS can find a schedule reaching the computation-time lower bound with high probability. Specifically, ${{\eta }_{n}}$ is no less than 98.70\%. Additionally, if a task-execution schedule does not reach the lower bound, it is still acceptable because the computation time is very close to the lower bound, i.e., ${{\gamma }_{n}}$ is no more than 6.48\%.

{\noindent{\bf{Case 2: power-of-two number of processors}}}

Note that we assume an arbitrary number of processors in Table II. We find from our experiments that for the case of $M$ equal to a power-of-two integer, all our tested task-execution schedules reach the lower bound and are optimal. For $n<7$, our algorithm is optimal in terms of the computation speed, as we tested all instances, and all the schedules reached the lower bound. For $n\ge 7$, since ${{\xi }_{0}}-{{\xi }_{1}}=0$, (\ref{eq_7}) in Section \ref{sec-II} can be further simplified as follows:
\begin{equation}\label{eq_27}
\begin{aligned}
&{{I}_{{\eta^L_{n}}}}
  \left( {{\xi }_{1}}+1,{{\xi }_{0}}-{{\xi }_{1}}+1 \right) \\
 & =\sum\limits_{\begin{smallmatrix}
 j={{\xi }_{1}}+1 \\ {{\xi }_{0}}={{\xi }_{1}}
\end{smallmatrix}}{\frac{\left( {{\xi }_{0}}+1 \right)!}{j!\left( {{\xi }_{0}}+1-j \right)!}\left({\eta^L_{n}}\right)^{j}{{\left( 1-{\eta^L_{n}} \right)}^{{{\xi }_{0}}+1-j}}} \\
 & =\left({\eta^L_{n}}\right)^{{{\xi }_{0}}+1} \\
\end{aligned}
.\end{equation}

With (\ref{eq_6}) and (\ref{eq_7}) in Section \ref{sec-II}, we have
\begin{equation}\label{eq_28}
{\eta^L_{n}}={{\left( 1-\alpha  \right)}^{\frac{1}{{{\xi }_{0}}+1}}}
.\end{equation}

Therefore, when $n\ge7$, we have ${{\eta }_{n}} \ge {{\left( 1-\alpha  \right)}^{\frac{1}{{{\xi }_{0}}+1}}}$, where $\alpha =0.95$ is the confidence level, and ${{\xi }_{0}}={{f}_{6}}=2598061$.
Additionally, we know that ${\gamma_n}=0$ from the definition of ${\gamma_{n,M}}$ in (\ref{eq_0B}).
The experimental results of ${{\eta }_{n}}$ and ${{\gamma }_{n}}$ are given in Table III.
\begin{table}[htbp]\label{table:3}
\caption{${{\eta }_{n}}$ and ${{\gamma }_{n}}$ for the scheduling problems in a ${{2}^{n}}$-point IFDMA FFT (the case of a power-of-two number of processors)}
\begin{center}
\begin{tabularx}{6.5cm}{p{1.0cm}<{\centering}p{3.1cm}<{\centering}p{1.4cm}<{\centering}}
\hline
$n$ & ${\eta _n}$ & ${\gamma _n}$ \\
\hline
\noalign{\smallskip}
\textbf{\textit{$3/4/5/6$}}&$1.0000$&$0.0000$\\
\noalign{\smallskip}
\textbf{\textit{$7/8/9/10$}}&$\ge \sqrt[{{\xi }_{0}}+1]{0.05}$&$0.0000$\\
\noalign{\smallskip}\hline
\end{tabularx}
\end{center}
\end{table}

The gap between $\sqrt[{{\xi }_{0}}+1]{0.05}$ and one is relatively small because ${{\xi }_{0}}$ is large enough, i.e., the lower bound of ${{\eta }_{n}}$ is relatively close to one. Thus, we can say that our algorithm has close-to-optimal performance when the number of processors is power-of-two.

\subsection{Benchmarking MPS-FFT Against Conventional FFT}
This subsection starts with a bin allocation example to illustrate how MPS-FFT saves computation time. After that, we benchmark the computation time of MSP-FFT against that of conventional FFT schemes.

Recall from the introduction that there are three conventional FFT schemes: serial FFT, pipelined FFT, and parallel FFT. Here we benchmark MPS-FFT against the first two schemes. Parallel FFT is not considered because it is impractically complex for modern OFDM communication systems with a large number of subcarriers \cite{ref_C1}. Pipelined FFT is the most widely used in FPGA/ASIC IP core for OFDM systems because the scheme has a good trade-off between hardware complexity and computation speed \cite{ref_C2}.

For a quick illustration, we first consider a 32-subcarrier compact IFDMA FFT with an example bin allocation written as ordered list $\{{{S}_{0}},{{S}_{1}},...{{S}_{5}}\}$, where ${{S}_{0}}=\{0,1...15\}$, ${{S}_{1}}=\{16,17...23\}$, ${{S}_{2}}=\{24,25,26,27\}$, ${{S}_{3}}=\{28,29\}$, ${{S}_{4}}=\{30\}$ and ${{S}_{5}}=\{31\}$.

\begin{figure}[htbp]
  \centering
  \includegraphics[width=3.0in]{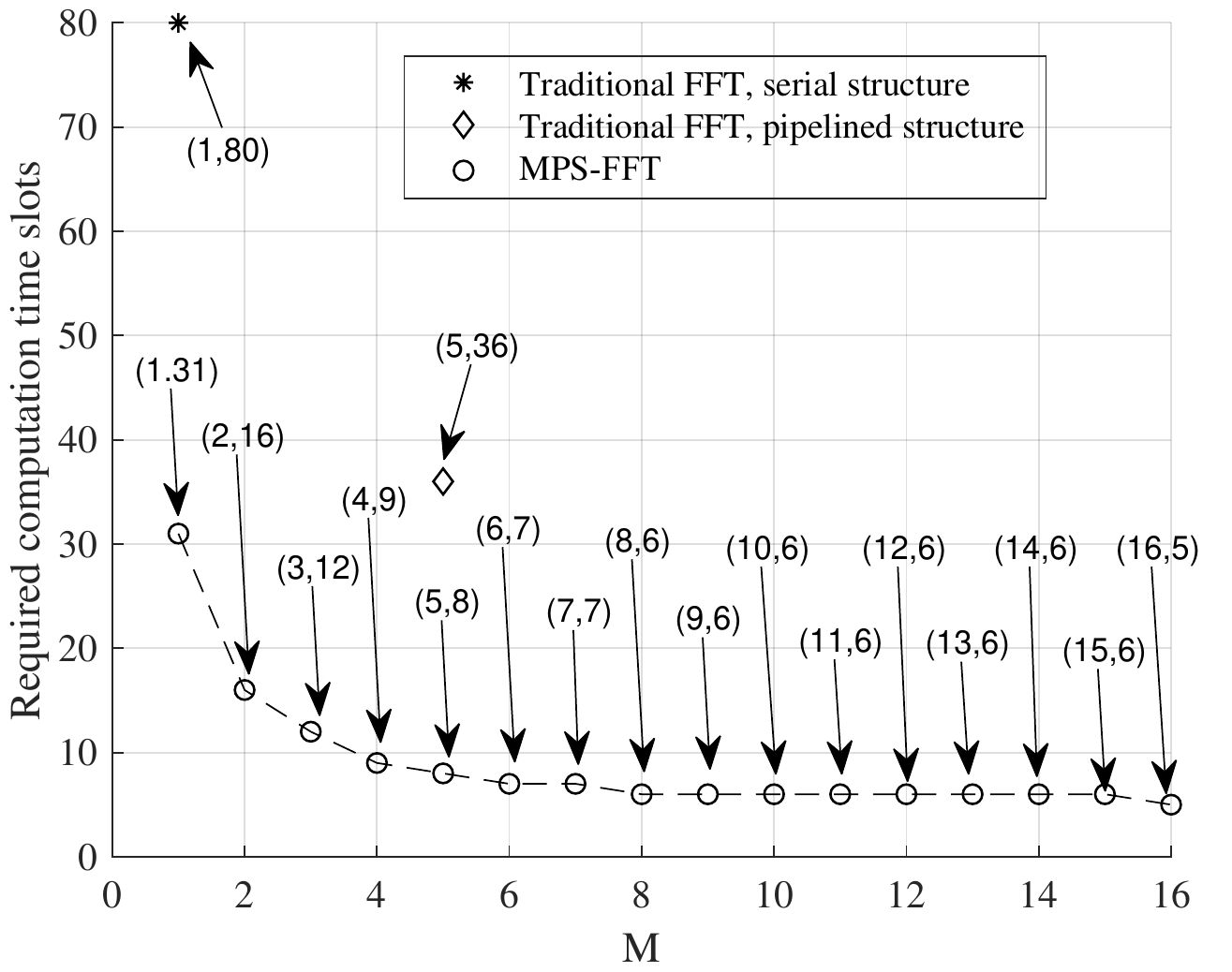}\\
  \caption{FFT computation times of different schemes. Point $(a,b)$ in the figure means that a scheduling scheme uses $a$ processors and consumes $b$ time slots for FFT computation with respect to the specific bin allocation $\{{{S}_{0}},{{S}_{1}},...{{S}_{5}}\}$.}
\label{fig:9}
\end{figure}

Fig. \ref{fig:9} shows the computation time versus the number of available processors for MPS-FFT. As shown, MPS-FFT only uses 31 time slots when $M=1$, significantly lower than the 80 time slots required by serial FFT, which also needs just one processor. MPS-FFT uses eight time slots when $M=5$, significantly lower than the 36 time slots required by pipelined FFT with five processors. MPS-FFT saves computation time in two ways, as explained below.

In Table IV, we present the FFT computation times of different schemes in a more general way. Here the sampling policy for large $n$ is consistent with that in the above subsection.
Note that the number of processors in conventional serial FFT and pipelined FFT are fixed to one and $\log_2 N$, respectively.
Therefore, we compare their computation times with MPS-FFT with one and $\log_2 N$ processors.
\begin{table}[htbp]\label{table:4}
\caption{Average number of time slots for different scheduling schemes}
\begin{center}
\begin{tabularx}{8.5cm}{p{0.1cm}<{\centering}p{1.55cm}<{\centering}p{1.65cm}<{\centering}p{1.55cm}<{\centering}p{1.65cm}<{\centering}}
\hline
\noalign{\smallskip}
$n$ & Serial FFT  & Pipelined FFT  & MPS-FFT & MPS-FFT \\
$\text{ }$ & $M=1$ & $M=log_2N$ & $M=1$ & $M=log_2N$ \\
\noalign{\smallskip}
\hline
\noalign{\smallskip}
\textbf{\textit{$3$}}&$12$&$9$&$8.50$&$3.70$\\
\noalign{\smallskip}
\textbf{\textit{$4$}}&$32$&$18$&$23.10$&$6.12$\\
\noalign{\smallskip}
\textbf{\textit{$5$}}&$80$&$35$&$62.18$&$12.84$\\
\noalign{\smallskip}
\textbf{\textit{$6$}}&$192$&$68$&$156.36$&$26.48$\\
\noalign{\smallskip}
\textbf{\textit{$7$}}&$448$&$133$&$376.72$&$54.25$\\
\noalign{\smallskip}
\textbf{\textit{$8$}}&$1024$&$262$&$881.41$&$110.61$\\
\noalign{\smallskip}
\textbf{\textit{$9$}}&$2304$&$519$&$2018.94$&$224.76$\\
\noalign{\smallskip}
\textbf{\textit{$10$}}&$5120$&$1032$&$4546.01$&$455.42$\\
\noalign{\smallskip}\hline
\end{tabularx}
\end{center}
\end{table}

\begin{figure}[htbp]
  \centering
  \includegraphics[width=3.3in]{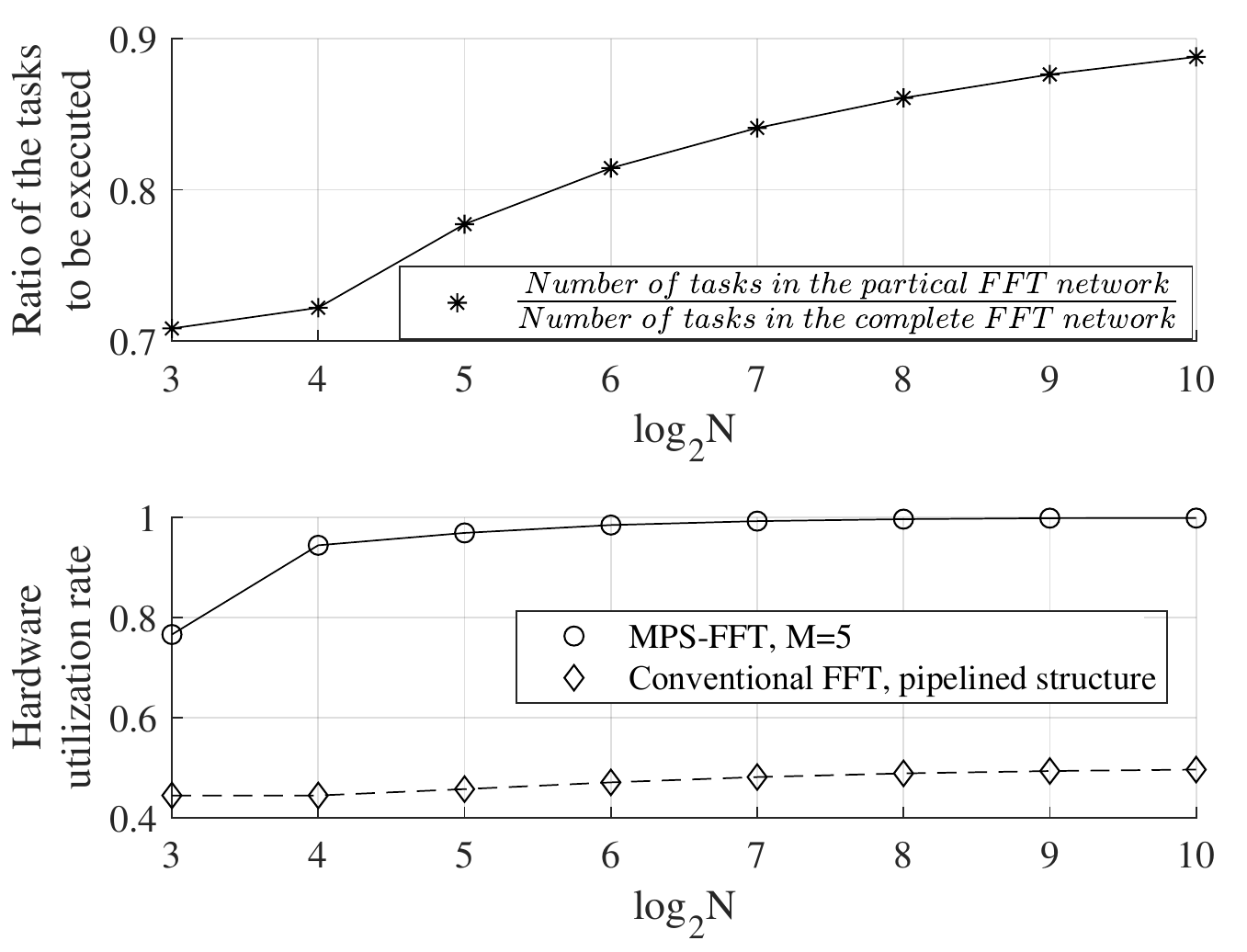}\\
  \caption{The general comparison results of the FFT computation in different schemes.}
\label{fig:10}
\end{figure}

Fig. \ref{fig:10} presents the data in Table IV.
Fig. \ref{fig:10}(a) plots the ratio of the number of the tasks in partial FFT (in MPS-FFT) and the number of tasks in complete FFT networks (for the conventional FFT computation).
In general, more than 11.21\% of the tasks can be exempted in the partial FFT network.

Fig. \ref{fig:10}(b) plots the hardware utilization rates of MPS-FFT and pipelined FFT, where the number of processors is set to be ${{\log }_{2}}N$.
As can be seen, the hardware utilization rate of the MPS-FFT is much higher than that of pipelined FFT.
Importantly, for FFT with 64 or more points, the hardware utilization rate of MPS-FFT is at least 98.42\%.
The rate increases when there are more subcarriers, reaching 99.82\% when there are 1024 subcarriers.

\begin{rem}
The pipelined FFT used for the benchmark is the classic and the most widely used FFT in practice, e.g., the same realization can be found in Xilinx FFT IP core design \cite{ref_D1}.
In this design, one processor is responsible for the tasks in one FFT stage, and the $\log_2 N$ processors work in a pipelined manner.
There have also been recent papers studying the hardware utilization of the pipelined FFT with a complete FFT network \cite{ref_D2,ref_D3,ref_D4}.
In these papers, a very high hardware utilization rate can be achieved through special hardware designs or scheduling schemes.
As shown in Fig. \ref{fig:10}(b), our MPS-FFT for the partial FFT network also has a very high hardware utilization rate, as these state-of-the-art pipelined FFT designs do in complete FFT networks.
\end{rem}

To conclude the experiment section, we summarize the merits of MPS-FFT as follows:
\begin{enumerate}[1)]
\item Given an arbitrary number of processors, MPS has at least a 98.70\% probability of finding the optimal task-execution schedule with the minimum computation time.
Furthermore, even if MPS does not find the optimal schedule, the computation time of the schedule found is at most 6.48\% higher than the lower bound on average.
\item When the number of processors is a power of two, MPS-FFT has a close-to-one probability of reaching the optimal computation time.
Quantitatively, the probability of reaching the lower bound is larger than $\sqrt[{{\xi }_{0}}+1]{0.05}$, where ${{\xi }_{0}}$ is more than 2.5 million in our experiment.
\item The complexity (in terms of the number of butterfly computations) of the partial FFT network in MPS-FFT is at least 11.21\% less than that of the complete FFT network in conventional FFT.
\item The processor utilization rate of MPS-FFT is at least 98.42\% when there are no less than 64 subcarriers.
When a large number of subcarriers are considered (e.g.,1024), the utilization rate can even go up to 99.82\%.
\end{enumerate}

\section{Conclusion}\label{sec-VI}
This paper studied the FFT implementation of an efficient IFDMA transceiver, referred to as compact IFDMA, put forth by a recent investigation. For compact IFDMA, not all butterfly computations inside the full FFT network need to be executed, and the necessary computations vary with different subcarrier allocations. When applied to IFDMA, conventional FFT implementation is resource-wasteful because it does not exploit this specific property of IFDMA signal processing.

This paper focused on FFT implementations tailored for IFDMA. We put forth a flexible heuristic algorithm to schedule the butterfly computations in IFDMA FFT, referred to as multi-priority scheduling (MPS). Compared with conventional FFT implementations, the FFT computation schedule obtained by MPS, referred to as MPS-FFT, has two advantages: 1) MPS-FFT reduces computation requirements. For example, for 1024-subcarrier IFDMA, MPS-FFT can bypass at least 11.21\% of the computation tasks in FFT. 2) MPS-FFT utilizes hardware efficiently. For example, for 1024-subcarrier IFDMA, the processor utilization rate in MPS-FFT is 99.82\%, much higher than the processor utilization rate in conventional pipeline FFT.

When the number of available processors is a power of two, MPS-FFT has near-optimal computation-time performance. Quantitatively, the probability that the schedule found by MPS is optimal is larger than $\sqrt[{{\xi }_{0}}+1]{0.05}$, where ${{\xi }_{0}}$ is more than 2.5 million in our experiment. Furthermore, MPS-FFT incurs less than 44.13\% of the computation time of the conventional pipelined FFT.

\appendices
\def\thesubsectiondis{{\arabic{subsection}}.}

\section{Supplementary Material for Section II-C}\label{sec-App2}
\setcounter{figure}{0}
\setcounter{table}{0}
\setcounter{equation}{0}
\renewcommand{\thefigure}{A\arabic{figure}}
\renewcommand{\thetable}{A\arabic{table}}
\renewcommand{\theequation}{A\arabic{equation}}
\subsection{Isomorphism of FFT Precedence Graphs}
In graph theory, isomorphism is an equivalence relation between graphs. The isomorphism of directed graphs ${{G}_{1}}\left( {{V}_{1}},{{E}_{1}} \right)$ and ${{G}_{2}}\left( {{V}_{2}},{{E}_{2}} \right)$ is defined as a bijection $h:{{V}_{1}}\to {{V}_{2}}$ between the vertex sets ${{V}_{1}}$ and ${{V}_{2}}$ , and a bijection $g:{{E}_{1}}\to {{E}_{2}}$ between the edge sets ${{E}_{1}}$ and ${{E}_{2}}$ such that there is a directed edge ${{e}_{1}}\in {{E}_{1}}$ from ${{v}_{1}}\in {{V}_{1}}$ to ${{v}_{2}}\in {{V}_{1}}$ if and only if there is a directed edge $g({{e}_{1}})\in {{E}_{2}}$ from $h({{v}_{1}})\in {{V}_{2}}$ to $h({{v}_{2}})\in {{V}_{2}}$. As a particular class of directed graphs with precedence relationships, the FFT precedence graphs also have isomorphism relationships.

Recall from Section II-A that we denote a bin allocation by an ordered list $\{{{S}_{0}},{{S}_{1}},...{{S}_{R-1}}\}$, where ${{S}_{r}}$ is the set of bins allocated to stream $r(r=0,1,...,R-1)$. We can further simplify the notation of bin allocation to $\left( {{s}_{0}},{{s}_{1}},...{{s}_{R-1}} \right)$, where ${{s}_{r}}$ is the number of bins stream $r$ has. For example, ${S_0} = \{ 0,1,2,3\} $ means that we allocate bin \#0 to bin \#3 to IFDMA stream 0, and ${{s}_{0}}=4$ indicates that we allocate the first four bins (which is also bin \#0 to bin \#3) to stream 0.

\begin{figure*}[htbp]
  \centering
  \includegraphics[width=7in]{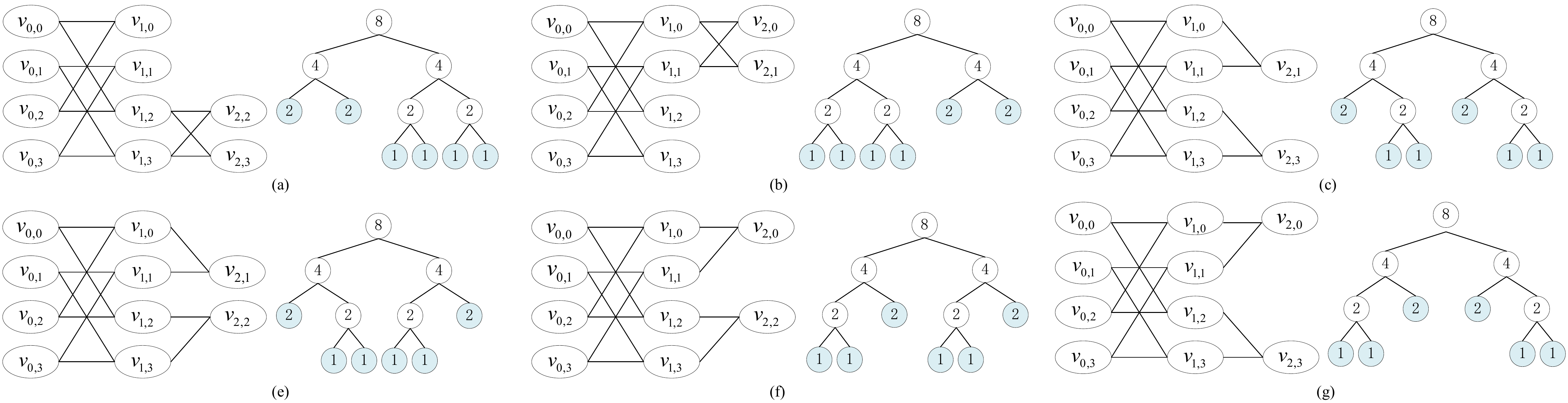}\\
  \caption{FFT precedence graphs and corresponding splitting trees for bin allocation  $\left( 2,2,1,1,1,1 \right)$, $\left( 1,1,1,1,2,2 \right)$, $\left( 2,1,1,2,1,1 \right)$, $\left( 2,1,1,1,1,2 \right)$, $\left( 1,1,2,1,1,2 \right)$, and $\left( 1,1,2,2,1,1 \right)$.}
\label{fig:11}
\end{figure*}

As discussed in Section II, we assume that the bins are fully allocated, i.e., $\sum\limits_{r=0}^{R-1}{{{s}_{r}}}={{2}^{n}}$. We also assume that the number of bins for each IFDMA stream is power-of-two. Accordingly, a bin allocation can be represented by the leaf nodes of a power-of-two splitting tree (abbreviated as splitting tree henceforth). For example, Fig. \ref{fig:11} presents the FFT precedence graphs and their corresponding splitting trees for bin allocations $\left( 2,2,1,1,1,1 \right)$, $\left( 1,1,1,1,2,2 \right)$, $\left( 2,1,1,2,1,1 \right)$, $\left( 2,1,1,1,1,2 \right)$, $\left( 1,1,2,1,1,2 \right)$, and $\left( 1,1,2,1,1,2 \right)$. Note that the node labeled ``8" in the splitting tree corresponds to the four vertexes at the first stage of the precedence graph; the two nodes labeled ``4" correspond to the upper two vertexes and the lower two vertexes of the second stage of the precedence graph, and so on. The leaf nodes do not correspond to any vertexes in the precedence graph, but they are the outputs of the childless vertexes in the precedence graphs (e.g., in Fig. \ref{fig:11}(a), the two leaf nodes labeled with ``2" are the outputs of vertexes ${{v}_{1,0}}$ and ${{v}_{1,1}}$). Thus, the number of stages in the splitting tree is one more than the number of stages in the precedence graph.

We could identify the isomorphism of two FFT precedence graphs by analyzing their splitting trees. If one splitting tree can be transformed to the other splitting tree by switching the positions of the two subtrees below a node, then the two FFT precedence graphs are isomorphic. Switching the positions of the two subtrees in the splitting tree as above corresponds to relabeling of a subset of vertexes of the precedence graph such that isomorphism applies. In Fig. \ref{fig:11}(a), for example, by switching the positions of the two subtrees below the root node ``8", we get the splitting tree in Fig. \ref{fig:11}(b). The precedence graph of Fig. \ref{fig:11}(a) becomes the precedence graph of Fig. \ref{fig:11}(b) with the following vertex relabeling: ${{v}_{1,0}}\to {{v}_{1,2}}$, ${{v}_{1,1}}\to {{v}_{1,3}}$, ${{v}_{1,2}}\to {{v}_{1,0}}$, ${{v}_{1,3}}\to {{v}_{1,1}}$, ${{v}_{2,2}}\to {{v}_{2,0}}$, ${{v}_{2,3}}\to {{v}_{2,1}}$.

Note that isomorphism is a transitive relationship: if A is isomorphic to B because by switching the positions of two subtrees below a node, the splitting tree of A can be transformed to that of B, and B is isomorphic to C because by switching the positions of two subtrees below another node, the splitting tree of B can be transformed to that of C, then A and C are isomorphic. For example, in Fig. \ref{fig:11}(c), if we switch the positions of the two subtrees under the right node ``4", we get the splitting tree in Fig. \ref{fig:11}(d), and if we further switch the positions of the two subtrees under the left node ``4", we get the splitting tree in Fig. \ref{fig:11}(e). From Fig. \ref{fig:11}(e), if we switch the positions of the two subtrees under the right node ``4", we get the splitting tree in Fig. \ref{fig:11}(f). Thus, Fig. \ref{fig:11}(c), (d), (e), and (f) are isomorphic.

We can say that $\left( 2,2,1,1,1,1 \right)$ and $\left( 1,1,1,1,22 \right)$ are isomorphic because their splitting trees are isomorphic. And we can also say that $\left( 2,1,1,2,1,1 \right)$, $\left( 2,1,1,1,1,2 \right)$, $\left( 1,1,2,1,1,2 \right)$, and $\left( 1,1,2,2,1,1 \right)$ are isomorphic for the same reason. The bin allocation $\left( 2,1,1,2,1,1 \right)$, on the other hand, is not isomorphic to $\left( 2,2,1,1,1,1 \right)$. Note, in particular, that although both $\left( 2,1,1,2,1,1 \right)$ and $\left( 2,2,1,1,1,1 \right)$ have two bins of size 2 and four bins of size 1, they are not isomorphic.

If the FFT precedence graphs ${{G}_{1}}$ and ${{G}_{2}}$ isomorphic, then if a scheduling algorithm can find the optimal schedule for ${{G}_{1}}$, then the algorithm should be able to find the optimal schedule for ${{G}_{2}}$. In fact, the performance of the scheduling algorithm does not change given the proper vertex relabeling. Thus, in studying the general performance of scheduling algorithms, we propose to focus on a set of non-isomorphic precedence graphs only so that isomorphic graphs are not over-represented if there are certain FFT graph structures with many isomorphic instances.

We denote the complete set of non-isomorphic FFT precedence graphs by ${{F}_{n}}$, and we denote the cardinality of ${{F}_{n}}$ by ${{f}_{n}}$. We refer to the bin allocations that lead to isomorphic FFT precedence graphs as isomorphic bin allocations. Among isomorphic bin allocations, we only select one for our experiments, as they make no difference in the optimality of a scheduling algorithm. We refer to the selected bin allocation as the non-isomorphic bin-allocation instance (non-isomorphic instance in short). We denote the complete set of non-isomorphic bin-allocation instances for ${{2}^{n}}$ subcarriers by ${{I}_{n}}$. The elements in ${{I}_{n}}$ map to the elements in ${{F}_{n}}$ in a one-to-one manner, and the cardinality of ${{I}_{n}}$ is also ${{f}_{n}}$.

Another question is how to select an instance from a group of isomorphic bin allocations for inclusion in ${{I}_{n}}$. We can randomly select any of the isomorphic bin allocations as it makes no difference in the optimality of a scheduling algorithm. However, we are interested in a deterministic test set. In this work, we select the ``right heavy" bin allocation. In Fig. \ref{fig:11}, subfigure (c), (d), (d), and (f) shows the splitting trees for isomorphic bin allocations $\left( 2,1,1,2,1,1 \right)$, $\left( 2,1,1,1,1,2 \right)$, $\left( 1,1,2,1,1,2 \right)$, and $\left( 1,1,2,2,1,1 \right)$, respectively. In each splitting tree, there are two subtrees that split a node of ``2" into two nodes of ``1". We select $\left( 2,1,1,2,1,1 \right)$ as the non-isomorphic bin allocation instance to include in ${{I}_{n}}$  because there are more leaf nodes to the right in its splitting tree.

\subsection{Recursion Function of $f_n$}
This subsection derives ${{f}_{n}}$, the cardinality of ${{F}_{n}}$, or equivalently, the cardinality of ${{I}_{n}}$.

Let $n=2$, it is easy to verify that ${{f}_{2}}=4$. The following table list the elements in ${{I}_{2}}$.
\begin{table}[htbp]
\caption{Non-isomorphic instances for the fully packaged compact IFDMA system with four subcarriers}
\begin{center}
\begin{tabularx}{5.5cm}{p{1cm}<{\centering}p{4.5cm}<{}}
\hline \noalign{\smallskip}
Index & Non-isomorphic Instance\\
\noalign{\smallskip}\hline\noalign{\smallskip}
\textbf{\textit{$0$}}&${I_2}(0) = (4)$\\
\noalign{\smallskip}
\textbf{\textit{$1$}}&${I_2}(1) = (2,2)$\\
\noalign{\smallskip}
\textbf{\textit{$2$}}&${I_2}(2) = \left( {2,1,1} \right)$\\
\noalign{\smallskip}
\textbf{\textit{$3$}}&${I_2}(3) = \left( {1,1,1,1} \right)$\\
\noalign{\smallskip}\hline
\end{tabularx}
\label{table:4}
\end{center}
\end{table}

\begin{figure}[htbp]
  \centering
  \includegraphics[width=3.25in]{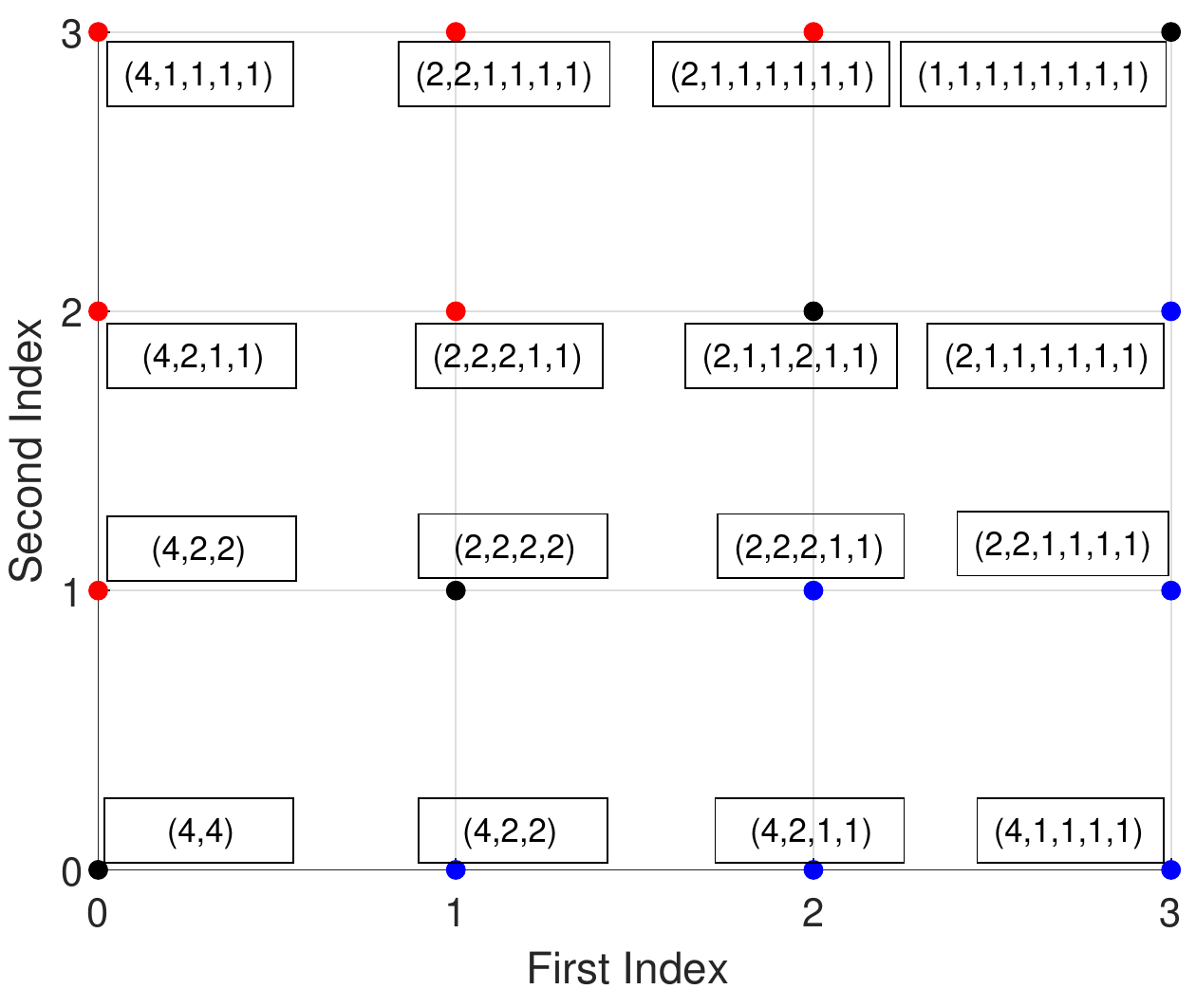}\\
  \caption{An illustration of how non-single-stream non-isomorphic instances for eight-subcarrier IFDMA is generated.}
\label{fig:12}
\end{figure}

We next explain how we obtain the non-isomorphic instances for $n=3$, i.e., the elements in ${{I}_{3}}$. If there is only one IFDMA stream using up all bins, the corresponding non-isomorphic instance is $(8)$. In the splitting tree, there is a root node ``8" only. If there is more than one IFDMA stream, then for the splitting tree, there are two subtrees under the root node ``8". Each subtree can be thought of as an instance in ${{I}_{2}}$. Thus, an instance in ${{I}_{3}}$, except $(8)$, can be thought of as a combination of two instances in ${{I}_{2}}$. We can treat ${{I}_{2}}$ as occupying a dimension on a two-dimensional grid, as shown in Fig. \ref{fig:12}. Each combination is one point on this two-dimensional space. Fig. \ref{fig:12} shows all possible combinations on the grid, where point $(i,j)$ represents the combination of  ${{I}_{2}}(i)$ and ${{I}_{2}}(j)$. Note that each combination in blue is isomorphic to a combination in red. Therefore, we only consider the combinations represented by points in red and black to remove isomorphism in ${{I}_{3}}$. In other words, we include only the points in the upper-left triangle of the grid.

Summing up the single-stream case and the non-single-stream cases, ${{f}_{3}}$ is then given by,
\begin{equation}
{f_3} = 1 + \left( {1 + 2 + ... + {f_2}} \right) = 1 + \frac{{{f_2}({f_2} + 1)}}{2} = 11
.\end{equation}

Table \ref{table:5} lists all the 11 non-isomorphic instances for $n=3$.
\begin{table}[htbp]
\caption{Non-isomorphic instances for the fully packaged compact IFDMA system with eight bins}
\begin{center}
\begin{tabularx}{5.5cm}{p{1cm}<{\centering}p{4.5cm}<{}}
\hline \noalign{\smallskip}
Index & Non-isomorphic Instance\\
\noalign{\smallskip}\hline\noalign{\smallskip}
\textbf{\textit{$0$}}&${{I}_{3}}(0)=(8)$ \\
\noalign{\smallskip}
\textbf{\textit{$1$}}&${{I}_{3}}(1)=(4,4)$ \\
\noalign{\smallskip}
\textbf{\textit{$2$}}&${{I}_{3}}(2)=(4,2,2)$\\
\noalign{\smallskip}
\textbf{\textit{$3$}}&${{I}_{3}}(3)=(4,2,1,1)$\\
\noalign{\smallskip}
\textbf{\textit{$4$}}&${{I}_{3}}(4)=(4,1,1,1,1)$\\
\noalign{\smallskip}
\textbf{\textit{$5$}}&${{I}_{3}}(5)=(2,2,2,2)$\\
\noalign{\smallskip}
\textbf{\textit{$6$}}&${{I}_{3}}(6)=(2,2,2,1,1)$\\
\noalign{\smallskip}
\textbf{\textit{$7$}}&${{I}_{3}}(7)=(2,2,1,1,1,1)$\\
\noalign{\smallskip}
\textbf{\textit{$8$}}&${{I}_{3}}(8)=(2,1,1,2,1,1)$\\
\noalign{\smallskip}
\textbf{\textit{$9$}}&${{I}_{3}}(9)=(2,1,1,1,1,1,1)$\\
\noalign{\smallskip}
\textbf{\textit{$10$}}&${{I}_{3}}(10)=(1,1,1,1,1,1,1,1)$\\
\noalign{\smallskip}\hline
\end{tabularx}
\label{table:5}
\end{center}
\end{table}

The same procedure can be used to find ${{I}_{n+1}}$ from ${{I}_{n}}$. Thus, in general, we have
\begin{equation}
{{f}_{n+1}}=1+\left( 1+2+...+{{f}_{n}} \right)=1+\frac{{{f}_{n}}({{f}_{n}}+1)}{2}
.\end{equation}

In this way, we obtain ${{f}_{n}}$ with various $n$. We list ${{f}_{1}}$ to ${{f}_{10}}$ in Table \ref{table:6}.
\begin{table}[htbp]
\caption{Number of non-isomorphic instances for $n=1,2,...,10$}
\begin{center}
\begin{tabularx}{8.0cm}{p{0.3cm}<{\centering}p{1.6cm}<{\centering}p{4.9cm}<{\centering}}
\hline \noalign{\smallskip}
$n$ & FFT Size ($N$) &Number of non-isomorphic instances ($f_n$)\\
\noalign{\smallskip}\hline\noalign{\smallskip}
\textbf{\textit{$1$}}&$2$&$2$\\
\noalign{\smallskip}
\textbf{\textit{$2$}}&$4$&$4$\\
\noalign{\smallskip}
\textbf{\textit{$3$}}&$8$&$11$\\
\noalign{\smallskip}
\textbf{\textit{$4$}}&$16$&$67$\\
\noalign{\smallskip}
\textbf{\textit{$5$}}&$32$&$2279$\\
\noalign{\smallskip}
\textbf{\textit{$6$}}&$64$&$2598061$\\
\noalign{\smallskip}
\textbf{\textit{$7$}}&$128$&$3.3750E+12$\\
\noalign{\smallskip}
\textbf{\textit{$8$}}&$256$&$5.6952E+24$\\
\noalign{\smallskip}
\textbf{\textit{$9$}}&$512$&$1.6218E+49$\\
\noalign{\smallskip}
\textbf{\textit{$10$}}&$1024$&$1.3151E+98$\\
\noalign{\smallskip}\hline
\end{tabularx}
\label{table:6}
\end{center}
\end{table}

\subsection{Test Set Generation Policy for $n\ge7$}
Note from Table \ref{table:6} that the number of non-isomorphic instances is not very large for $n\le6$. Therefore, when doing experiments to obtain  $\left( {{\eta }_{n,M}},{{\gamma }_{n,M}} \right)$, we use ${{I}_{n}}$, the complete set of the non-isomorphic instance as the test set. Specifically, we can store all the instances in ${{I}_{n}}$ in our program (e.g., as an array), and subject all the instances to the scheduling algorithm in a one-by-one manner in our experiments. The statistics $\left( {{\eta }_{n,M}},{{\gamma }_{n,M}} \right)$ can then be obtained from the experiments.

For $n\ge7$, however, ${{f}_{n}}$ is so large that it is impractical to test all the instances in ${{I}_{n}}$. Furthermore, even storing them in the program could be problematic. Therefore, we consider a subset of ${{I}_{n}}$ as the test set and characterize $\left( {{\eta }_{n,M}},{{\gamma }_{n,M}} \right)$ statistically. We denote the subset by $I_{n}^{'}$, and the cardinality of $I_{n}^{'}$ is written as $f_{n}^{'}$. In our experiments, we set $f_{n}^{'}={{\xi }_{0}}=f_6$ (i.e., we fix the cardinality of $I_{n}^{'}$ to a constant ${{\xi }_{0}}<{{f}_{n}}$) for all $n\ge7$

Next, we explain how we obtain the non-single-stream elements $I_{n+1}^{'}$ with a recursive operation. We set the initial condition of the recursive analysis as $I_{6}^{'}={{I}_{6}}$ and $f_{6}^{'}={{f}_{6}}$. In this way, in the two-dimensional grid representing non-single-stream instances in the case of $n+1=7$, each dimension has the complete set ${{I}_{6}}$. We randomly select $f_{n}^{'}={{\xi }_{0}}$ points from the upper left triangle of the two-dimensional grid ${{I}_{6}}\times {{I}_{6}}$ plus the bin allocation $({{2}^{7}})$ to form $I_{7}^{'}$. The ${{f}_{7}}$ instances are selected with equal probability to form the $f_{7}^{'}$ instances in $I_{7}^{'}$, the mechanism of which will be explained shortly.

For the two-dimensional grid ${{I}_{7}}\times {{I}_{7}}$ that represents the $n+1=8$ case, ${{I}_{7}}$ is so large that it is impractical to have the complete set of ${{I}_{7}}$ in each dimension. Therefore, we use $I_{7}^{'}$, the  random subset described above in place of ${{I}_{7}}$ in each of the two dimensions. We then randomly sample $f_{n}^{'}={{\xi }_{0}}$ points from the upper left triangle of the $I_{7}^{'}\times I_{7}^{'}$ grid to form $I_{8}^{'}$. In the case of $n+1=9$, we randomly sample from the upper left triangle of the $I_{8}^{'}\times I_{8}^{'}$ grid to form  $I_{9}^{'}$, and so on and so forth.

Next, we present how we perform the random selection of instances from the upper left triangle of $I_{n}^{'}\times I_{n}^{'}$ to form $I_{n+1}^{'}$, where $n+1\ge 7$, to ensure that the instances are generated with equal probability.

First, we consider the instance where there is only one IFDMA stream, $({{2}^{n+1}})$ . With a probability of ${1}/{{{f}_{n+1}}}$ (note that this is ${1}/{{{f}_{n+1}}}$ rather than ${1}/{f_{_{n+1}}^{'}}$), we generate the non-isomorphic instance $({{2}^{n+1}})$.

If the generation in the first step fails, we go to the non-single-stream instances and generate an element in $I_{n+1}^{'}$ by combining $I_{n}^{'}(i)$ and $I_{n}^{'}(j)$. In essence, we are trying to generate a duple $(i,j)$ with constrains 1) $0\le i\le f_{_{n}}^{'}-1$ and 2) $i\le j\le f_{_{n}}^{'}-1$ with equal probability. We denote the probability by a constant $c$, and we have
\begin{equation}
{{P}_{i,j}}={{P}_{i}}\cdot {{P}_{\left. j \right|i}}={{P}_{i}}\cdot \frac{1}{f_{n}^{'}-i}=c
,\end{equation}
where ${{P}_{i}}$ is the probability of a specific $i$ value is selected between $0$ and $f_{_{n}}^{'}-1$. After $i$ is selected, ${{P}_{\left. j \right|i}}$ denotes the probability of a specific $j$ is selected between $i$ and $f_{_{n}}^{'}-1$.

We know that there are altogether ${f_{n}^{'}\left( f_{n}^{'}+1 \right)}/{2}$ instances in the upper left triangle of $I_{n}^{'}\times I_{n}^{'}$ grid that form the non-single-stream elements in $I_{n+1}^{'}$. Normalizing the sum of the selection probabilities all instances to 1, we have
\begin{equation}
\sum\limits_{i,j\ge i}{{{P}_{i,j}}}=\sum\limits_{i,j\ge i}{{{P}_{i}}\cdot {{P}_{\left. j \right|i}}}=\frac{f_{n}^{'}\left( f_{n}^{'}+1 \right)}{2}c=1
.\end{equation}

Hence,
\begin{equation}
c=\frac{2}{f_{n}^{'}\left( f_{n}^{'}+1 \right)}
.\end{equation}
and ${{P}_{i}}$ can be therefore written as,
\begin{equation}
{{P}_{i}}=c\left( f_{n}^{'}-i \right)=\frac{2\left( f_{n}^{'}-i \right)}{f_{n}^{'}\left( f_{n}^{'}+1 \right)},i=0,1,...,f_{n}^{'}-1
.\end{equation}

After an integer between $0$ and $f_{_{n}}^{'}-1$ is assigned to $i$, with equal probability we select $j$ between $i$ and $f_{_{n}}^{'}-1$. We have
\begin{equation}
{{P}_{\left. j \right|i}}=\frac{1}{f_{n}^{'}-i},j=i,i+1,...,f_{n}^{'}-1
.\end{equation}

From the above equations, we can see that
\begin{equation}
\sum\limits_{i=0}^{f_{n}^{'}-1}{{{P}_{i}}}=1
\end{equation}
and
\begin{equation}
\sum\limits_{\begin{smallmatrix}
 i=0,...,f_{n}^{'}-1
 \text{, }
 j=i,...,f_{n}^{'}-1
\end{smallmatrix}}{{{P}_{\left. j \right|i}}}=1
\end{equation}
are satisfied.

We now summarize how we generate $(i,j)$ as follows:
\begin{enumerate}[1)]
\item We first generate an integer $i$ between $0$ and $f_{_{n}}^{'}-1$ probability ${{P}_{i}}={2\left( f_{n}^{'}-i \right)}/{f_{n}^{'}\left( f_{n}^{'}+1 \right)}$ (note: that means a larger integer has a smaller probability of being assigned to $i$).
\item We then generate an integer $j$ between $i$ and $f_{_{n}}^{'}-1$ with equal probability ${{P}_{j|i}}={1}/{f_{n}^{'}-i}$.
\end{enumerate}

After we get $(i,j)$, we can obtain an element of $I_{n+1}^{'}$ by combining $I_{n}^{'}(i)$ and $I_{n}^{'}(j)$.

\section{The Full Butterfly Structure Inside An FFT Precedence Graph}\label{sec-App4}
\begin{cor}[Butterfly Structure of the Precedence Graph]\label{cor:1}
Consider a vertex ${{v}_{i,{{j}_{1}}}}$ with two children ${{v}_{i+1,{{j}_{C1}}}}$ and ${{v}_{i+1,{{j}_{C2}}}}$, and another vertex ${{v}_{i,{{j}_{2}}}}$ with two children ${{v}_{i+1,{{j}_{C3}}}}$ and ${{v}_{i+1,{{j}_{C4}}}}$, in an FFT precedence graph. If ${{j}_{C1}}={{j}_{C3}}$, then ${{j}_{C2}}={{j}_{C4}}$ (i.e., for two-child parents, if they share a child, they must also share the other child, giving rise to a butterfly relationship in the precedence graph).
\end{cor}
\begin{NewProof}
First, suppose that ${{j}_{1}}\bmod \left( {{2}^{{{\log }_{2}}N-i-1}} \right)<{{2}^{{{\log }_{2}}N-i-2}}$, part 1) of Property 4 states that ${{j}_{C1}}={{j}_{1}}$ and ${{j}_{C2}}={{j}_{1}}+{{2}^{{{\log }_{2}}N-i-2}}$.
\begin{enumerate}[1)]
\item For ${{v}_{i{+}1,{{j}_{C1}}}}$, since ${{j}_{C1}}\bmod \left( {{2}^{{{\log }_{2}}N-(i+1)}} \right)<{{2}^{{{\log }_{2}}N-(i+1)-1}}$, part 1) of Property 3 states that its two parents are ${{v}_{i,{{j}_{C1}}}}$ and ${{v}_{i,{{j}_{C1}}+{{2}^{{{\log }_{2}}N-i-2}}}}$.
\item For ${{v}_{i{+}1,{{j}_{C2}}}}$, since ${{j}_{C2}}\bmod \left( {{2}^{{{\log }_{2}}N-(i+1)-1}} \right)\ge {{2}^{{{\log }_{2}}N-(i+1)-2}}$, part 2) of Property 3 states that its two parents are ${{v}_{i,{{j}_{C2}}-{{2}^{{{\log }_{2}}N-i-2}}}}$ and ${{v}_{i,{{j}_{C2}}}}$.
\end{enumerate}

Note in particular that ${{j}_{1}}={{j}_{C1}}={{j}_{C2}}-{{2}^{{{\log }_{2}}N-i-2}}$ and ${{j}_{C2}}={{j}_{C1}}+{{2}^{{{\log }_{2}}N-i-2}}$, i.e., except the common parent ${{v}_{i,{{j}_{1}}}}$, ${{v}_{i+1,{{j}_{C1}}}}$ and ${{v}_{i+1,{{j}_{C2}}}}$ have another common parent ${{v}_{i,{{j}_{2}}}}$, where ${{j}_{2}}={{j}_{C2}}={{j}_{C1}}+{{2}^{{{\log }_{2}}N-i-2}}$.

Next, suppose that ${{j}_{1}}\bmod \left( {{2}^{{{\log }_{2}}N-i-1}} \right)\ge {{2}^{{{\log }_{2}}N-i-2}}$, by similar reasoning, we can arrive at the same conclusion that two parents of ${{v}_{i+1,{{j}_{C1}}}}$ and ${{v}_{i+1,{{j}_{C2}}}}$ are the same, i.e., their common parents are  ${{v}_{i,{{j}_{1}}}}$ and ${{v}_{i,{{j}_{2}}}}$, where ${{j}_{1}}={{j}_{C1}}={{j}_{C2}}+{{2}^{{{\log }_{2}}N-i-2}}$ and ${{j}_{2}}={{j}_{C2}}={{j}_{C1}}-{{2}^{{{\log }_{2}}N-i-2}}$.

Finally, we conclude that if ${{v}_{i,{{j}_{1}}}}$ has two children ${{v}_{i+1,{{j}_{C1}}}}$ and ${{v}_{i+1,{{j}_{C2}}}}$, there must be another vertex at stage $i$ (denoted by ${{v}_{i,{{j}_{2}}}}$) sharing the same two children with ${{v}_{i,{{j}_{1}}}}$. In particular, the relationship between the two parents and two children is of a butterfly structure.
\end{NewProof}

\section{A Serialization Scheme to Map a Priority Vector to A Priority Scalar}\label{sec-App5}
\setcounter{figure}{0}
\setcounter{table}{0}
\setcounter{equation}{0}
\renewcommand{\thefigure}{C\arabic{figure}}
\renewcommand{\thetable}{C\arabic{table}}
\renewcommand{\theequation}{C\arabic{equation}}
In the following, we sketch a possible serialization process for that purpose. Note that we first focus on the general MPS algorithm framework where $H$ can be any positive integer. Then, we consider the $H=4$ case for our MPS algorithm in this paper.

Suppose that ${{P}_{h}}\in \{0,1,...,p_{h}^{\max }-1\}$ and that $q=\underset{h}{\mathop{\max }}\,p_{h}^{\max }$. Let ${{P}_{h}}({{v}_{i,j}})$ denote the ${{P}_{h}}$ of vertex ${{v}_{i,j}}$. Then, the priority scalar of vertex ${{v}_{i,j}}$ is given by the following serialization equation:
\begin{equation}\label{eq_12}
P({v_{i,j}}) = \sum\limits_{h = 1}^H {{\alpha _h}{P_h}({v_{i,j}})}
,\end{equation}
where ${{\alpha }_{h}}$ is the weighted coefficient of ${{P}_{h}}$ given by,
\begin{equation}\label{eq_13}
{\alpha _h}{\kern 1pt}  = {q^{H - h}}
.\end{equation}

In other words, with (\ref{eq_12}) and (\ref{eq_13}), we are mapping a priority vector to an $H$-digit $q$-ary scalar number. It is easy to see that the priority order is preserved with this mapping. We emphasize that this is not the only serialization possible, nor is it the most ``economical" in terms of the size of the scalar: it is just a simple alternative to understand the basic mechanism. Instead of (\ref{eq_13}), we could also have an alternative coefficient definition as follows:
\begin{equation}\label{eq_14}
{{\alpha }_{h}}=1+\prod\limits_{i=1}^{h-1}{p_{i}^{\max }}
.\end{equation}

Let us continue with alternative (\ref{eq_13}) in the following. For the $H=4$ case in our MPS algorithm, we know the ranges of ${{P}_{h}}(h=0,1,2,3)$ are,
\begin{equation}\label{eq_15}
\left\{ \begin{array}{l}
{P_4} \in \left\{ {0,1,...,{N \mathord{\left/
 {\vphantom {N 2}} \right.
 \kern-\nulldelimiterspace} 2} - 1} \right\}\\
{P_3} \in \left\{ {0,1,2} \right\}\\
{P_2} \in \left\{ {0,1,2} \right\}\\
{P_1} \in \left\{ {0,1,...,\log_2 N - 1} \right\}
\end{array} \right.
.\end{equation}

For IFDMA system with $N\ge4$, we have $q{=}{N}/{2}\;$. Hence, the overall priority can be expressed as,
\begin{equation}\label{eq_16}
P({{v}_{i,j}})=\sum\limits_{h=1}^{H=4}{{{\left( \frac{N}{2} \right)}^{H-h}}{{P}_{h}}({{v}_{i,j}})}
.\end{equation}

\section{Pseudocode of the Scheduling Algorithm}\label{sec-App3}
We denote the set of executable vertexes by $\Omega $. Algorithm 1 is the main function of the MPS algorithm. Algorithm 2 is the initialization subfunction called at the beginning of Algorithm 1.  Algorithm 3 is the priority scalar calculation subfunction described by eq. (\ref{eq_16}), which is also called by Algorithm 1.

\begin{algorithm}[htbp]
  \caption{MPS Algorithm ($H=4$)}
  \begin{algorithmic}[1]
    \Require
        IFDMA bin allocation $\{{{S}_{0}},{{S}_{1}},...{{S}_{R-1}}\}$ and $M$
    \Ensure
       Task schedule ${\bf{X}} =\left\{ {{\chi }_{0}},{{\chi }_{0}},...{{\chi }_{T-1}} \right\}$ and $T$
       \State $t\leftarrow 0$ \Comment{Algorithm initialization}
       \State $\left( {\Omega,\text{ }G(V,E),\text{ }\left\langle {{P_1},{P_2},{P_3},{P_4}} \right\rangle } \right) \leftarrow$ initialize func.
       \While {$G(V,E)\ne \varnothing$}
           \State $u\leftarrow 0$
           \While {$\left(u<M \text{ and }\Omega \ne \varnothing\right)$} \Comment{Vertex selection}
                \State ${v_{i,j}} \leftarrow \mathop {\arg }_{v \in \Omega } \max P\left( v \right)$
                \State ${\chi _t}{\kern 1pt}.{\kern 1pt}{\kern 1pt}push{\kern 1pt} ({v_{i,j}})$
                \State $\Omega{\kern 1pt}.{\kern 1pt} {\kern 1pt}pop{\kern 1pt}  ({v_{i,j}})$
                \If{ ${{v}_{i,j}}$ has companion ${{v}_{{{i}^{\mathbf{'}}},{{j}^{\mathbf{'}}}}}$ and ${{v}_{{{i}^{\mathbf{'}}},{{j}^{\mathbf{'}}}}} \in \Omega$ } \Comment{If a vertex is selected, raise $P_2$ of its companion}
                    \State ${{P}_{2}}({{v}_{{{i}^{\mathbf{'}}},{{j}^{\mathbf{'}}}}})\leftarrow 2$
                    \State $P({{v}_{{{i}^{'}},{{j}^{'}}}})\leftarrow$ priority scalar calculation func.
                \EndIf
                \State $u\leftarrow u+1$
           \EndWhile \Comment{End of vertex selection in one time slot}
           \State $V{\kern 1pt}.{\kern 1pt} {\kern 1pt} pop{\kern 1pt}(elements {\kern 1pt} {\kern 1pt} in {\kern 1pt} {\kern 1pt} {\chi _t})$
           \State $E{\kern 1pt}.{\kern 1pt} {\kern 1pt} pop{\kern 1pt}( output{\kern 1pt} {\kern 1pt} edges{\kern 1pt} {\kern 1pt} of{\kern 1pt} {\kern 1pt} the {\kern 1pt} {\kern 1pt} elements {\kern 1pt} {\kern 1pt} in {\kern 1pt} {\kern 1pt} {\chi _t})$
           \State $\Omega {\kern 1pt}.{\kern 1pt} {\kern 1pt} push{\kern 1pt} (new{\kern 1pt} {\kern 1pt} executable{\kern 1pt} {\kern 1pt} vertexes)$
           \State $t \leftarrow t + 1$
       \EndWhile \Comment{End of a time slot}
  \end{algorithmic}
\end{algorithm}

\begin{algorithm}[htbp]
  \caption{Priority Scalar Calculation}
  \begin{algorithmic}[1]
    \Require
        IFDMA bin allocation $\{{{S}_{0}},{{S}_{1}},...{{S}_{R-1}}\}$
    \Ensure
        Initialized
        (1)FFT precedence graph;
        (2)executable vertexes set $\Omega $;
        (3)priority vectors $\left\langle {{P_1},{P_2},{P_3},{P_4}} \right\rangle $
    \State $G(V,E) \leftarrow \text{build}{\kern 1pt}{\kern 1pt} \text{FFT}{\kern 1pt}{\kern 1pt} \text{precedence}{\kern 1pt}{\kern 1pt} \text{graph} {\kern 1pt}{\kern 1pt} \text{with}{\{ {S_0}...{S_{R - 1}}\} }$ \Comment{See Section II-A on how to build the graph}
    \For{$i = log_2N-1$; $i\ge0$; $i\text{-}\text{-}$ }
        \For {$j = 0$; $j\le log_2N-1$; $j\text{++}$ }
            \If{ ${{v}_{i,j}}\in {{V}}$ }
                \State $K({v_{i,j}}) \leftarrow {\kern 1pt}{\kern 1pt} \text{child} {\kern 1pt}{\kern 1pt} \text{of}{\kern 1pt}{\kern 1pt} v_{i,j}$
                \State ${k_{i,j}} \leftarrow {\kern 1pt}{\kern 1pt} \text{cardinality} {\kern 1pt}{\kern 1pt} \text{of} {\kern 1pt}{\kern 1pt} K({v_{i,j}}) $
                \If{$k_{i,j}=0$}
                    \State ${P_1}({v_{i,j}}) \leftarrow 0$
                    \State ${P_2}({v_{i,j}}) \leftarrow 0$
                \Else
                    \State ${P_1}({v_{i,j}}) \leftarrow 1 + \mathop {\max }\limits_{{v_{i + 1,x}} \in K({v_{i,j}})} {P_1}({v_{i + 1,x}})$
                    \State ${P_2}({v_{i,j}}) \leftarrow 1$
                \EndIf
                \State ${P_3}({v_{i,j}}) \leftarrow {k_{i,j}}$
                \State ${P_4}({v_{i,j}}) \leftarrow {N \mathord{\left/{\vphantom {N 2}} \right.\kern-\nulldelimiterspace} 2} - 1 - j$
            \EndIf  
            \If{ $i=0$ }
                \State $\Omega {\kern 1pt}.{\kern 1pt} {\kern 1pt} push{\kern 1pt} (v_{i,j})$
            \EndIf  
        \EndFor 
    \EndFor
  \end{algorithmic}
\end{algorithm}

\begin{algorithm}[htbp]
  \caption{Priority Scalar Calculation}
  \begin{algorithmic}[1]
    \Require
        Priority vector ${\left\langle {{P_1}({v_{i,j}}),{P_2}({v_{i,j}}),{P_2}({v_{i,j}}),{P_3}({v_{i,j}})} \right\rangle }$
    \Ensure
        Priority scalar $P({v_{i,j}})$
    \State $P({v_{i,j}}) = \sum\limits_{h = 1}^{H = 4} {{{\left( {{N \mathord{\left/ {\vphantom {N 2}} \right. \kern-\nulldelimiterspace} 2}} \right)}^{H - h}}{P_h}({v_{i,j}})} $ \Comment{see eq. (\ref{eq_16})}
  \end{algorithmic}
\end{algorithm}

\bibliographystyle{IEEEtran}
\bibliography{00A.Arxiv_Full_Version_Baseline}

\begin{thebibliography}{10}
\providecommand{\url}[1]{#1}
\csname url@samestyle\endcsname
\providecommand{\newblock}{\relax}
\providecommand{\bibinfo}[2]{#2}
\providecommand{\BIBentrySTDinterwordspacing}{\spaceskip=0pt\relax}
\providecommand{\BIBentryALTinterwordstretchfactor}{4}
\providecommand{\BIBentryALTinterwordspacing}{\spaceskip=\fontdimen2\font plus
\BIBentryALTinterwordstretchfactor\fontdimen3\font minus
  \fontdimen4\font\relax}
\providecommand{\BIBforeignlanguage}[2]{{%
\expandafter\ifx\csname l@#1\endcsname\relax
\typeout{** WARNING: IEEEtran.bst: No hyphenation pattern has been}%
\typeout{** loaded for the language `#1'. Using the pattern for}%
\typeout{** the default language instead.}%
\else
\language=\csname l@#1\endcsname
\fi
#2}}
\providecommand{\BIBdecl}{\relax}
\BIBdecl

\bibitem{Ref_B1}
S.~Han, Y.~C. Liang, B.~H. Soong, and S.~Li, ``Dynamic broadband spectrum
  refarming for {OFDMA} cellular systems,'' \emph{IEEE Trans. Wireless
  Commun.}, vol.~15, no.~9, pp. 6203--6214, 2016.

\bibitem{Ref_B2}
H.~G. Myung, J.~Lim, and D.~J. Goodman, ``Single carrier {FDMA} for uplink
  wireless transmission,'' \emph{IEEE Veh. Technol. Mag.}, vol.~1, no.~3, pp.
  30--38, 2006.

\bibitem{Ref_B3}
Y.~Zhu and K.~B. Letaief, ``{CFO} estimation and compensation in {SC-IFDMA}
  systems,'' \emph{IEEE Trans. Wireless Commun.}, vol.~9, no.~10, pp.
  3200--3213, 2010.

\bibitem{ref_D5}
Y.~Du, J.~Chen, Y.~Lei, and X.~Hao, ``Performance analysis of nonlinear spatial
  modulation multiple-input multiple-output systems,'' \emph{Digital Signal
  Process.}, vol. 115, p. 103064, 2021.

\bibitem{Ref_B4}
S.~Zhang, S.~Xu, G.~Y. Li, and E.~Ayanoglu, ``First 20 years of green radios,''
  \emph{IEEE Trans. Green Commun. and Netw.}, vol.~4, no.~1, pp. 1--15, 2020.

\bibitem{ref_1}
S.~C. Liew and Y.~Shao, ``New transceiver designs for interleaved
  frequency-division multiple access,'' \emph{IEEE Trans. Wireless Commun.},
  vol.~19, no.~12, pp. 7765--7778, 2020.

\bibitem{ref_2}
Y.~Shao and S.~C. Liew, ``Flexible subcarrier allocation for interleaved
  frequency division multiple access,'' \emph{IEEE Trans. Wireless Commun.},
  vol.~19, no.~11, pp. 7139--7152, 2020.

\bibitem{ref_B5}
C.~F. Hsiao, Y.~Chen, and C.~Y. Lee, ``A generalized mixed-radix algorithm for
  memory-based {FFT} processors,'' \emph{IEEE Trans. Circuits Syst. II Express
  Briefs}, vol.~57, no.~1, pp. 26--30, 2010.

\bibitem{Ref_B6}
M.~Garrido, ``A survey on pipelined {FFT} hardware architectures,'' \emph{J.
  Signal Process. Syst.}, pp. 1--20, 2021.

\bibitem{Ref_B7}
Y.~T. Ma, ``A {VLSI}-oriented parallel {FFT} algorithm,'' \emph{IEEE Trans.
  Signal Process.}, vol.~44, no.~2, pp. 445--448, 1996.

\bibitem{Ref_A1}
R.~M. Karp, ``Reducibility among combinatorial problems,'' in \emph{Complex.
  Comput.}\hskip 1em plus 0.5em minus 0.4em\relax Springer, 1972, pp. 85--103.

\bibitem{Ref_A2}
J.~K. Lenstra and A.~Rinnooy~Kan, ``Complexity of scheduling under precedence
  constraints,'' \emph{Oper. Res.}, vol.~26, no.~1, pp. 22--35, 1978.

\bibitem{Ref_A3}
H.~Jia~Wei and H.~T. Kung, ``{I/O} complexity: The red-blue pebble game,'' in
  \emph{ACM STOC}, 1981, pp. 326--333.

\bibitem{Ref_A5}
M.~Bahtat, S.~Belkouch, P.~Elleaume, and P.~Le~Gall, ``Instruction scheduling
  heuristic for an efficient {FFT} in {VLIW} processors with balanced resource
  usage,'' \emph{EURASIP J. Adv. Signal Process.}, vol. 2016, no.~1, pp. 1--21,
  2016.

\bibitem{Ref_A10}
R.~B.~Ramakrishna, ``Iterative modulo scheduling,'' [Online]. Available:
  \url{https://www.hpl.hp.com/techreports/94/HPL-94-115.pdf}.

\bibitem{Ref_A6}
M.~Frigo, ``A fast fourier transform compiler,'' in \emph{ACM PLDI}, 1999, pp.
  169--180.

\bibitem{Ref_A7}
M.~Frigo and S.~G. Johnson, ``{FFTW}: An adaptive software architecture for the
  {FFT},'' in \emph{IEEE ICASSP}, vol.~3.\hskip 1em plus 0.5em minus
  0.4em\relax IEEE, 1998, pp. 1381--1384.

\bibitem{Ref_A8}
A.~Ali and L.~Johnsson, ``{UHFFT}: A high performance {DFT} framework,'' 2006.

\bibitem{Ref_A9}
M.~Puschel, J.~M. Moura, J.~R. Johnson, D.~Padua, M.~M. Veloso, B.~W. Singer,
  J.~Xiong, F.~Franchetti, A.~Gacic, Y.~Voronenko \emph{et~al.}, ``{SPIRAL}:
  Code generation for {DSP} transforms,'' \emph{Proc. IEEE}, vol.~93, no.~2,
  pp. 232--275, 2005.

\bibitem{BetaHandBook}
M.~Abramowitz, I.~A. Stegun, and R.~H. Romer, ``Handbook of mathematical
  functions with formulas, graphs, and mathematical tables,'' 1988.

\bibitem{CentralLimitTheorem}
R.~Durrett, \emph{Probability: theory and examples}.\hskip 1em plus 0.5em minus
  0.4em\relax Cambridge university press, 2019.

\bibitem{ref_C1}
L.~Kronsj\"{o}, \emph{Computational Complexity of Sequential and Parallel
  Algorithms}.\hskip 1em plus 0.5em minus 0.4em\relax USA: John Wiley \& Sons,
  Inc., 1986.

\bibitem{ref_C2}
I.~Cho, T.~Patyk, D.~Guevorkian, J.~Takala, and S.~Bhattacharyya, ``Pipelined
  {FFT} for wireless communications supporting 128--2048/1536-point
  transforms,'' in \emph{2013 IEEE GlobalSIP}.\hskip 1em plus 0.5em minus
  0.4em\relax IEEE, 2013, pp. 1242--1245.

\bibitem{ref_D1}
Xilinx, ``Fast fourier transform v9.1 {LogiCORE IP} product guide,'' [Online].
  Available:
  \url{https://www.xilinx.com/support/documentation/ip_documentation/xfft/v9_1/pg109-xfft.pdf}.

\bibitem{ref_D2}
Y.~N. Chang and K.~Parhi, ``An efficient pipelined {FFT} architecture,''
  \emph{IEEE Trans. Circuits Syst. II, Analog Digit. Signal Process.}, vol.~50,
  no.~6, pp. 322--325, 2003.

\bibitem{ref_D3}
L.~Yang, K.~Zhang, H.~Liu, J.~Huang, and S.~Huang, ``An efficient locally
  pipelined {FFT} processor,'' \emph{IEEE Trans. Circuits Syst. II Express
  Briefs}, vol.~53, no.~7, pp. 585--589, 2006.

\bibitem{ref_D4}
M.~Garrido, S.~J. Huang, S.~G. Chen, and O.~Gustafsson, ``The serial commutator
  {FFT},'' \emph{IEEE Trans. Circuits Syst. II Express Briefs}, vol.~63,
  no.~10, pp. 974--978, 2016.

\end{thebibliography}
\end{document}